\RequirePackage{fix-cm}
\documentclass[twocolumn]{svjour3}          
\smartqed  
\usepackage[sort&compress,numbers]{natbib}
\usepackage{graphicx}
\usepackage{amsmath}
\usepackage{multirow}
\usepackage{mathtools, cuted}
\usepackage{enumitem}
\usepackage{float}
\usepackage{algorithm} 
\usepackage{algpseudocode} 
\newlist{alphalist}{enumerate}{1}
\setlist[alphalist,1]{label=\textbf{\alph*.}}
\usepackage{caption}
\usepackage{subcaption}
\usepackage{multirow}
\newcommand*{\affaddr}[1]{#1} 
\newcommand*{\affmark}[1][*]{\textsuperscript{#1}}
\usepackage{fancyhdr}
\begin{document}

\title{EHRP : An effective hybrid routing protocol to compromise between energy consumption and delay in WSNs 
}


\author{%
Noureddine Moussa {\protect\affmark[1]$^{*}$} \and Zakaria Hamidi-Alaoui \affmark[1]  \and Abdelbaki El Belrhiti El Alaoui \affmark[1] 
}
\authorrunning{N. Moussa et al.}

\institute{
              $^{*}$Noureddine Moussa \\
              \email{n.moussa@edu.umi.ac.ma}   
                \\[8pt]                        
               Zakaria Hamidi-Alaoui \at 
             \email{z.hamidialaoui@edu.umi.ac.ma} 
              \\[8pt]                        
               Abdelbaki El Belrhiti El Alaloui \at 
             \email{a.elbelrhiti@fs-umi.ac.ma} 
              \\[8pt]
               \affaddr{\affmark[1]Computer Networks and Systems Laboratory, Faculty of Sciences, Moulay Ismail University of Meknes, PB 11201 Zitoune, 50000 Meknes, Morocco} \\
              \\[8pt]        
                }

\date{Received: date / Accepted: date}

\maketitle

\begin{abstract}
Sink mobility is seen as a successful strategy to resolve the hotspot problem in Wireless Sensor Network (WSN). Mobile sinks roam in the network and collect data from special nodes such as Cluster Heads (CH) by means of short-range communications which improves the energy efficiency. Numerous mobile sink based routing protocols have been proposed, however, they incur high delays especially in large scale networks where the mobile sink has to travel for a long distance to collect data from CHs and consequently they failed to ensure a tradeoff between energy efficiency and delay. To resolve this issue, we propose in this paper an Effective Hybrid Routing Protocol termed as EHRP. The main aim of this protocol is to combine between single-hop and multi-hop routing. Indeed, when the mobile sink arrives at a cluster it collects its data while the other distant CHs continue to send their data using our proposed improved Ant Colony Optimization (ACO) algorithm to avoid the waiting-time. The existing ACO algorithms use in the distance heuristic which is not practical in real world and fail to consider relevant statistic information of energy (e.g., minimum energy, average energy) in path selection which leads to unbalanced  energy consumption in the network. To address these issues, the proposed routing algorithm employs the Received Signal Strength Indicator (RSSI) and statistic information of energy to consume energy efficiently and decrease the probability of sending failure. The performance of the proposed
routing protocol is tested and compared with those of the relevant routing protocols. The simulation results show that, in comparison with its counterparts, EHRP succeeds to minimize energy consumption and delay as well as enhancing the packet delivery ratio. 
\end{abstract}

\section{Introduction}
\label{}

A WSN is made up of low-cost, energy-autonomous devices capable of monitoring physical or environmental conditions (temperature, humidity, noise, vibration, pressure, movement, pollution, etc.), of performing certain calculations, and of collaborating to transmit their data using single hop or multi-hop routing via wireless links to a final user via a unique node called sink. The energy resource of the sensor node is generally a battery. However, in most sensor network applications, the nodes are deployed in hostile environments and it is impractical to recharge or replace their batteries. Therefore, the overall lifetime of the network depends completely on that of the battery. Among the many existing solutions for effectively minimizing energy consumption in WSNs, one of the most recognized is the hierarchical organization of the network into clusters \cite{fanian_rafsanjani_2019}. However, in multi-hop cluster-based
WSNs, the CHs near the sink consume more energy than the rest of nodes since each CH gathers, transmits
the data packets from its own cluster members, and also relays the collected data of other CHs. Accordingly, the CHs
near the sink deplete their energy rapidly, which results in energy holes and premature network death \cite{wang_gao_yin_li_kim_2018}. This problem is known as the hotspot problem. One of the prominent solutions proposed to overcome this issue is the sink mobility \cite{singh_kumar_2019}. The mobile sink travels in the network to collect data through short communications distances. Hence, the load on CHs is alleviated owing to the fact that the sink visits each CH in turn to collect its data via 1-hop communication scheme. As a result, significant energy is saved and the network lifetime is increased. Mobile sink methods are devised into two main categories i.e. single mobile sink and multiple mobile sinks. Several works  \cite{mottaghi_zahabi_2015,nazir_hasbullah_2010,wang_cao_sherratt_park_2017} have been done in the first category but they suffer from the delay incurred during data collection. To overcome this issue, the authors in  \cite{krishnan_yun_jung_2019,vahabi_eslaminejad_dashti_2019,singh_kumar_singh_2016} have proposed the use of many mobile sinks. Indeed, in a large area, collecting
data from an individual sensor node by mobile
sinks is energy-efficient but it takes a long time that
means large delay. In a nutshell, these methods failed to ensure a compromise between energy efficiency and delay. Therefore, the aim of this paper is to realize this compromise in order to enhance the network performance.

In this paper, we allow the clustering to be done only once to avoid energy overhead and save significant amount of energy \cite{moussa}. In parallel with this, we propose the use of multiple mobile sinks in order to alleviate the hotspot problem. So in each round, a mobile sink moves to a given cluster and the other CHs in other clusters send their data to this sink using a multi-hop routing algorithm. The literature has indicated that the ACO algorithm is very useful for data routing \cite{s2019}. It is clear to see that significant efforts \cite{ref2015,r1,r2,r3,r4,r5,r6} have been made to improve energy efficiency and balancing using ACO algorithm. However, when applying the ACO to WSN routing optimization, still there are the following shortage that decreases its performance and precludes its practical application. Most researchers consider the distance between two nodes as the efficient factor in energy consumption, but this is not
valid in all circumstances because if there is an obstacle between nodes, the obtained signal's intensity (power) will be small. Also, these works ignore the impacts of energy statistics (e.g., average energy, minimum energy) which brings unbalanced  energy consumption.  In the proposed work, the heuristic function is improved using RSSI and the path is selected based on energy statistics and path length.

In summary, the proposed work allows 1) energy efficiency via CH selection in a round-robin method and round by round sink data gathering technique and the proposed improved ACO algorithm as well as (2) lower latency such that individual CHs do not wait until the coming of the sink but they still send their data to the sink.

The main contributions of this paper are :

\begin{itemize}

\item We devise an improved heuristic function which considers RSSI for energy efficiency and reliable data transmission purposes. Also, the path is selected based on statistic information of energy (e.g. average energy, minimum energy).

\item Existing mobile sink based routing protocols failed to ensure tradeoff between energy efficiency and delay especially in large scale networks. Thus, to fix this issue, we propose to combine single hop and multi-hop schemes within a clustered and multiple mobile sinks based WSN.

\item We perform extensive simulations on the proposed protocol and compare its results with those of two recent and relevant protocols \cite{krishnan_yun_jung_2019,vahabi_eslaminejad_dashti_2019}. The simulation results show that our proposed
protocol performs better than these protocols with respect
to different main metrics : energy consumption, network lifetime, average delay, and packet delivery
ratio.
\end{itemize}

The remainder of this paper is organized as follows. In section~\ref{rel}, we present the related works. Section~\ref{prop} detailed our proposed routing protocol. Section~\ref{ther} gives a theoretical analysis of delay and lifetime. The performance evaluation is given in Section~\ref{perf}. Section~\ref{conc} concludes and gives future work.

\section{Related works}
\label{rel}

Many algorithms for WSNs have been discussed in the literature, especially for addressing the hotspot problem. some recent of these algorithms use unequal clustering strategy \cite{ref1,ref2,ref3,ref4,ref5}. Their principle, is to have smaller sized clusters towards the sink and ensure availability of more CHs near the sink to share the high data-forwarding load. Because the remaining energy of CHs decreases more likely over time compared to that of the other nodes, it become logical to reduce the size of clusters to conserve their energy. However, the cluster sizes in the mentioned algorithms do not adapt to the change in residual energy of the CHs. To fix this issue, the authors in \cite{ref6} have proposed Distributed Unequal Cluster-based Routing (DUCR) protocol. But this protocol does not control the number of created CHs, which requires massive control overhead consuming a lot of energy \cite{ref7}. A major drawback of unequal clustering is that the data message takes a long time to reach the sink, which is not adequate for delay sensitive WSN applications. For that, many recent applications introduced mobile sinks as a solution. Thus, instead of forwarding data of sensor nodes in a multi-hop fashion to a static sink, mobile sinks travel near to all CHs to collect data directly through a single hop, which reduces both the number of hops and the latency for delivering data from sensor nodes to the sink. Existing literature reveals two approaches of using mobile sink, which are adopted in several recent works: either using a single mobile sink \cite{ref8,ref9,ref10,ref2015}  or using multiple of them \cite{ref11,ref12,krishnan_yun_jung_2019,vahabi_eslaminejad_dashti_2019}.

\begin{itemize}
\item \textbf{Single mobile sink based works}
\end{itemize}

The authors in \cite{ref8} proposed a Hybrid Method, called HM-ACOPSO, which combines ACO and Particle Swarm Optimization (PSO) to schedule an efficient moving path for the mobile sink. Indeed, the network is divided into several clusters each one is managed by a CH. To protect the weak nodes, the communication range of CHs can be adjusted according to the remaining energy, and anchor nodes can be merged to save the sojourn time. Then, the mobile sink travels along a predefined trajectory to traverse all the data collection sojourn points. For that, an efficient traveling loop is scheduled for the mobile sink using both methods ACO and PSO. 

In \cite{ref9}, the authors study the multi-hop data forwarding problem and suggests a novel tree-clustering scheme named Energy Efficient Tree Clustering (EETC) which aims at minimizing the energy consumption and extending the network lifetime while maintaining an optimized tour delay of the mobile sink. EETC employs a heuristic clustering method called Optimal Generation of Clusters (OGENCL) in the clustering phase. In this work, the number of relay hops between a CH and cluster members has been restricted to ensure network load balancing. 

A data collection strategy based on ACO with mobile sink is suggested for industrial WSNs in \cite{ref10}. In the aim to reduce the number of sensor nodes visited by the mobile sink and minimize the traversed path, the election of rendezvous nodes based on entropy weight technique is proposed according to some parameters such as relative remaining energy, nodes density, and the degree of distribution uniformity. In addition, an ACO algorithm is suggested to get the optimal traversed path for the mobile sink.

Another novel routing algorithm \cite{ref2015} combines the features of ACO, clustering and sink mobility techniques. This algorithm adopts the ideas of Low-Energy Adaptive Clustering Hierarchy (LEACH) in the CH election and enhances the CHs rotation process by considering the residual energy of nodes. To interact and collect data from each CH via short-range communications, the ACO algorithm is used to find the optimal route for the mobile sink to travel to every CH in the network.

As the above-mentioned algorithms use a single mobile sink, they may likely suffer from high delays in the moment of data collection. To alleviate this issue, many works in the literature proposed new algorithms that are based on multiple mobile sinks.

\begin{itemize}
\item \textbf{Multiple mobile sink based works}
\end{itemize}

In \cite{ref11}, the authors proposed a trajectory scheduling approach based on coverage rate for multiple mobile sinks for large scale WSNs. The authors combine an enhanced particle swarm optimization and mutation operator to seek for sojourn points with optimal coverage rate. To prepare the moving path for several mobile sinks, the genetic algorithm is then incorporated.

In \cite{ref12}, the authors suggested an unequal clustering approach using fuzzy logic by dividing the respective region into several smaller areas and by using multiple mobile sinks. This reduces the distances between sensor nodes with regard to a sink movement path. As a consequence, cluster sizes are decreased. Accordingly, such a decline in cluster sizes and the smart route selection for mobile sinks could remove the problem of energy hole.

The authors in \cite{krishnan_yun_jung_2019} proposed an Enhanced Clustering and ACO-based Multiple Mobile Sinks (ECMMS) approach to improve the data gathering efficiency and network lifespan of WSNs. A modified LEACH-based clustering technique is also implemented here so that the CHs are divided into a finite number of disjoint subsets; each of which is assigned to one mobile sink. Then, every mobile sink begins its travel from its initial position, traverses all the CHs in its assigned subset, and ends the tour at the same initial position. The traversed path is determined by ACO.

In \cite{vahabi_eslaminejad_dashti_2019}, the authors suggest the Integration of Geographic and Hierarchical Routing (IoGHR) with mobile sinks for energy conservation in WSNs. The IoGHR 's network model consists of virtual grids where each one contains a CH. A mobile sink moves back and forth through the virtual grids flowing a predetermined movement path. Each time the mobile sink reaches near to a CH in each virtual grid, it automatically starts gathering data from that CH.

The above algorithms based on ACO are interested in finding shortest path and ignore the current  node energy which leads to premature death of some nodes and affects the network lifespan. It is interesting to mention that using multiple mobile sinks can mitigate the hotspot problem and accordingly enhance the energy efficiency as well as reduce the delay in data collection to a some extent by visiting the maximum of data suppliers but in large scale networks it is impossible to visit all the CHs at the same time and then some CHs are visited late. Therefore, it is required to compromise between delay and energy efficiency. To provide a suitable solution to all these challenges, we present in the following the proposed solution.

\section{The proposed work}
\label{prop}

In this section, we present the energy model, the virtual grid creation method, the clustering algorithm, the movement strategy for the mobile sink and the data routing algorithm.

\subsection{Energy model}

In our scenario, we consider the most used energy model for data transmission and reception. Hence, the energy consumption consists of (i) the necessary energy to turn ON the radio transceiver to send the \textit{l}-bits message, which is a function of message length only and does not depend on the communication distance and (ii) the necessary energy to amplify the signal to reach the intended receiver. This energy depends on both the message length (in bits) and the communication distance. The energy consumption for sending \textit{l}-bits of data over a distance \textit{d} can be computed as expressed in equation 1, where $E_{elec}$ is the energy required to transmit data over the wireless medium, $\varepsilon_{tworayamp}$ and $\varepsilon_{freespaceamp}$ are respectively the amplification coefficients for the two-ray and free space communication models, and $d_{0}$ is the crossover distance. 
\begin{equation}
 \label{eqq}
 E_{TX} ( l, d ) = \begin{cases}
l * E_{elec} + q * \varepsilon_{tworayamp}*d^{4}, d \geq d_{0}  \\ 
l * E_{elec} + q * \varepsilon_{freespaceamp}*d^{2}, d < d_{0}
\end{cases}
\end{equation}
 
To receive \textit{l} bits of data, the required energy at the receiver is given by equation 2.
\begin{equation}
E_{RX} (l) = E_{elec} * l 
\end{equation}

\begin{figure}[h]
\centering
\includegraphics[width=8.5cm]{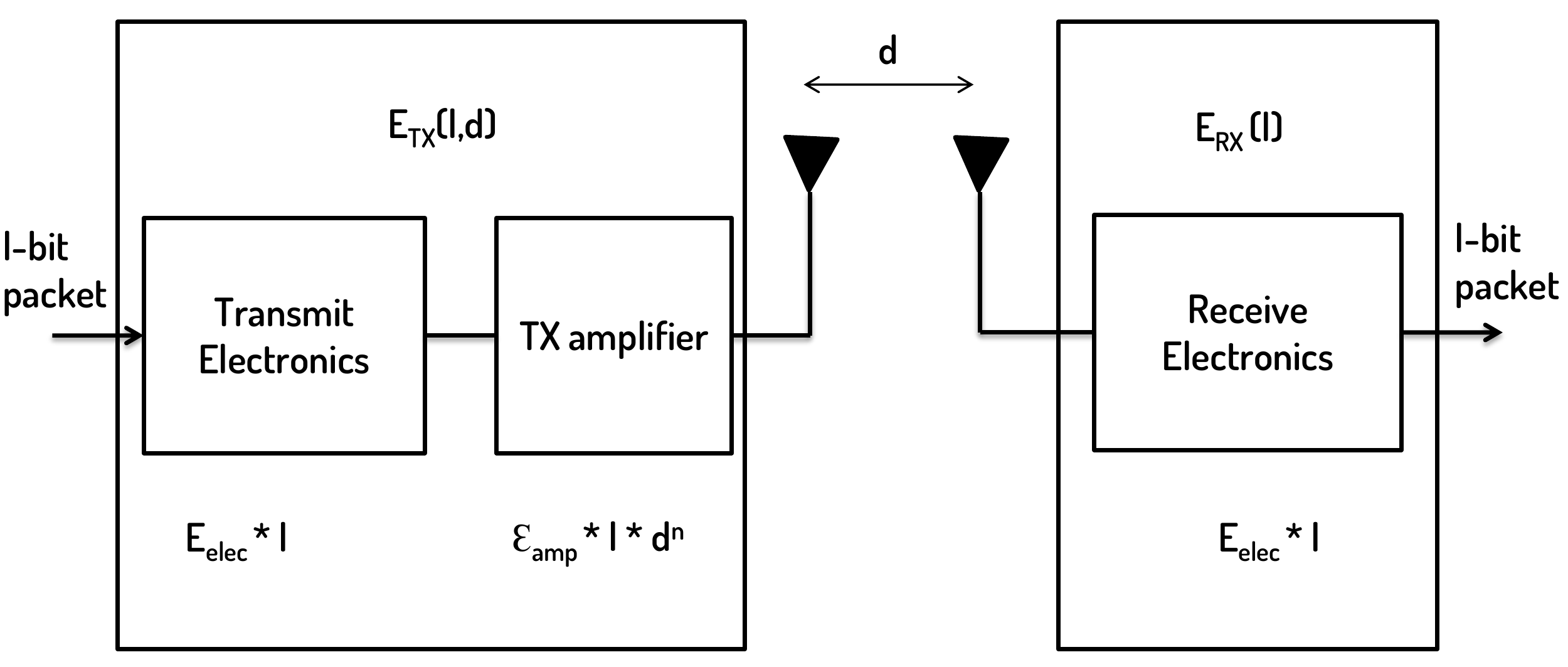}
\caption{Radio energy dissipation model \cite{Ref463}.}
\label{enmo}
\end{figure}

The radio energy dissipation model is depicted in figure~\ref{enmo}.

\subsection{Virtual grid creation}

The network contains a partition subdivided into many virtual grids of equal size. Every virtual grid consists of four sensor nodes and a CH  \cite{vahabi_eslaminejad_dashti_2019}. The CH collects data from other sensor nodes in the virtual grid and transmits it to the sink. The number of virtual grids can be estimated in the current proposed method by equation~\ref{eq1}, while the number of CHs can be determined by equation~\ref{eq2}, where \textit{N} represents the number of sensor nodes in the network.
\begin{equation}
\label{eq1}
    N_{virtualgrids} = \frac{N}{5}
\end{equation}
\begin{equation}
\label{eq2}
    N_{CHs} = N_{virtualgrids}
\end{equation}

At the network start, the mobile sink forwards its initial position in each virtual grid to sensor nodes. Since the mobile sink movement path is predefined, each sensor node can then estimate the current position of the mobile sink. In this way, the mobile sink sends its location just once, avoiding then excessive control overhead. Therefore, a great amount of energy could be saved.

\subsection{Clustering algorithm}

The CH selection policy employed in this work is conceived in such a way to guarantee the powerful node energetically to be elected as CH every round. Therefore, a node with the highest energy in each virtual grid is selected as the CH node. The role of CH is rotated fairly and based on energy among the actual CH and its members. The CH selection is carried out once until the network ends its service in order to save
energy. A novel and dynamic method has been used for this purpose. The basic idea is to rotate the CH role among all the cluster members. At each round, every sensor node within a cluster should know the CH for the next round. This is ensured by giving an order to each node based on its energy. After several rounds, all its cluster members inform their current CH about their residual energy. This allows to choose the potential Ch for the next rounds and therefore avoid frequent CH selection so that the energy consumption is minimized. It is known that in large scale networks, some sensor nodes may not be attached to a CH. This is the problem of connectivity and the proposed clustering method takes care of these nodes called unconnected nodes. It assigns them to a CH based on overhearing the messages circulated in network between connected nodes and their CHs. Also, the proposed clustering algorithm is fault tolerant as it allows to faulty cluster members to join a backup CH in case of failure of their primary CH in the aim to ensure the normal functioning of the WSN.

\subsection{Mobile sink mechanism}

\begin{figure}[h]
\centering
\includegraphics[width=8.5cm]{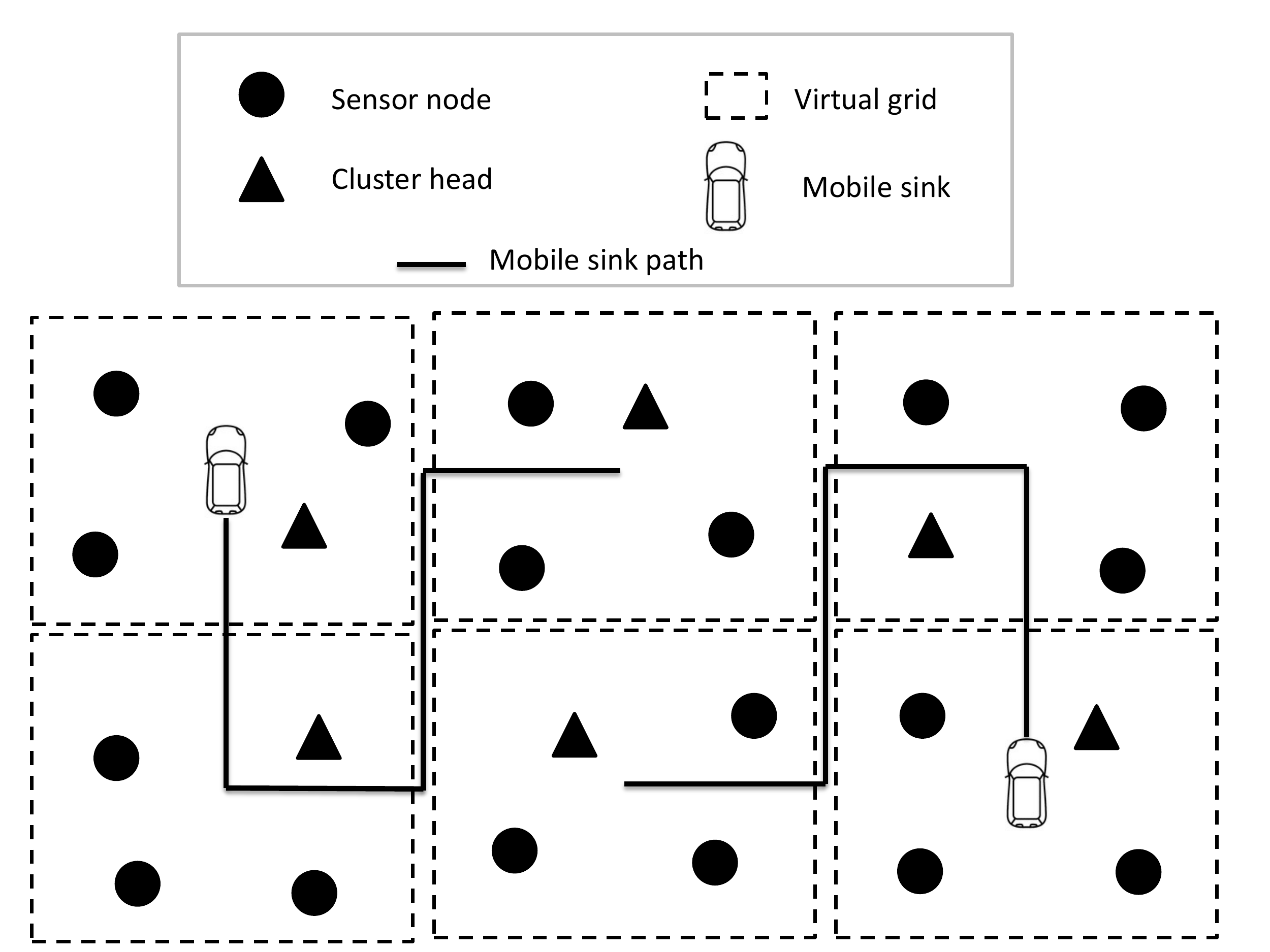}
\caption{Network model.}
\label{fig2}
\end{figure}

In order to solve the hotspot problem, the sink mobility technology is introduced in many WSNs routing protocols. Generally, there exist three types of mobile sink movement out of which we find random, controlled, and predefined path. In random, mobile
sink moves randomly in the network while in controlled mobile sink moves based on certain conditions like the energy. In
the predefined path, the mobile sink travels following a predetermined path. A predefined trajectory for the mobile sink is the most suitable trajectory
for WSNs, where there is no need for overhead resulting from message exchange to update the route trajectory which saves a lot of energy. In this study, as in figure~\ref{fig2} we use a predefined mobility for the
sink. Within a partition, the sink moves back and forth
through virtual grids. The sink node is located at
fixed position for the first time and on the basis of a time schedule, the mobile sink roams in the network. This ensures that CHs in each virtual grid or other virtual grids know the
time schedule and the sink location. The same process is done using multiple mobile sinks. Thus, some virtual grids are allocated for each mobile sink.
That is to say, each mobile sink belongs to a number of virtual grids such that to avoid data conflict. 

\subsection{Data routing algorithm}

The main concern of our work is to combine the advantages of mobile sink, clustering and the ACO based multi-hop routing algorithm to ensure a compromise between energy efficiency and delay. To have a better energy efficiency, multiple mobile sinks and clustering the network in a efficient way may present good results. However, the delay question still arises since the mobile sink has to traverse CHs to collect their data in turn role which incurs important delays. So, the basic idea in this paper is to perform routing in one single hop and multi-hop manners. In each round the sink moves to a certain grid to gather data of the CH while other CHs in other grids send their data through multi-hop to the sink using our improved ACO based algorithm. 

\begin{figure}[h]
\centering
\includegraphics[width=8.5cm]{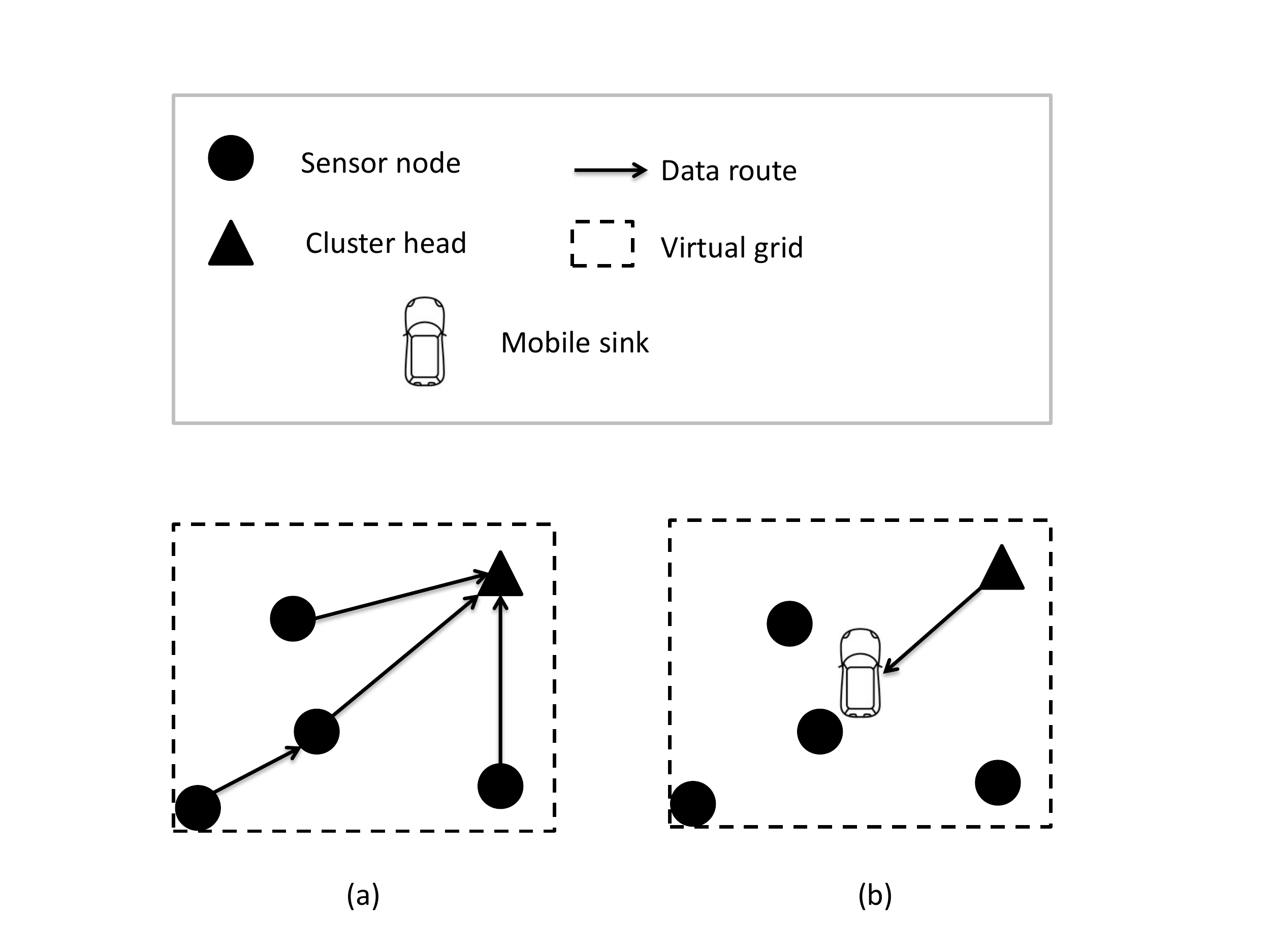}
\caption{Data transmitting.}
\label{fig1}
\end{figure}

The single hop communication mode is explained as follows. Sensor nodes send their data to their respective CH in the virtual grid. In turn, CHs gather data of cluster members, compress them and send them in the specific time based on periodical movement of the sink. The sink travels back and forth through virtual grids in the partition and when it is near each CH it collects data from CHs. This means that the sink collects data using single hop by means of close communications which further saves a high amount of energy in the network. Figure~\ref{fig1}a represents data forwarding from sensor nodes to the CH in the virtual grid, and figure~\ref{fig1}b illustrates the data routing from CH to the mobile sink.

In the same time, faraway CHs from the sink when it is in a virtual grid for a round, send their data using multi-hop routing method to minimize delays and avoid waiting for the sending of their data. The routing is ensured using an improved ACO algorithm. The routing method is explained in the following. 

Originally proposed for finding optimal paths in a graph, the ACO algorithm is inspired by the behavior of ants looking for a path between their colony and a food source. The original idea has since branched out to solve a larger class of problems, and several algorithms have emerged, inspired by various aspects of ant behavior. ACO can produce better results.
Therefore, in this work we apply the ACO algorithm to find
the shortest path to route data. 

The probabilistic formula used in traditional ACO presented in equation~\ref{AC}. It is the probability $p_{ij}^k (t)$ of the $k$th ant moving from the node \textit{i} to the node \textit{j}. 

\begin{equation}
\label{AC}
p_{ij}^k (t) = \left\{ \begin{array}{l}
\frac{  \tau_{ij}^\alpha (t)  \cdot \eta_{ij}^\beta }
{\sum\limits_{k \in allowed_{k}}{\tau_{ik}^\alpha (t) \cdot \eta_{ik} ^\beta (t) } } 
 {\kern 11pt} , if j \in allowed_{k} \\
 0 {\kern 93pt}   , Otherwise \\
 \end{array} \right.\
\end{equation}
where $\tau_{ij(t)}$ is the pheromone trail amount on arc (\textit{i},\textit{j}) at time \textit{t}. $\eta_{ij}$ represents the link visibility, $allowed_{k}$ denotes the nodes that are not visited by ant \textit{k}. $\alpha$ and $\beta$ are constant parameters used to adjust the pheromones' impact. 

Despite the strengths of the traditional ACO algorithm and its versions, it has some disadvantages like (1) The search time for these algorithms is long, and the convergence speed of the algorithms is slow. (2) The algorithm can fall into local optimal solution rather than a
global solution. (3) The use of distance  which is not practical in real world.  These issues have been fixed by considering the RSSI factor to select the next node so as to enhance the global search ability, avoid to fall in a local optimal solution, enhance the convergence rate and save energy. Therefore, the network delay is thus decreased. To this end, $\eta_{ij}$ is improved by using the maximum sum of RSSI between the current node to the next node and the next node to the destination node, thus:

\begin{equation}
\label{eq3}
\eta_{ij} =  { max [RSSI(i, j ) + RSSI( j,m)] } 
\end{equation}
where \textit{RSSI(i,j)} is the RSSI between node \textit{i} and next node \textit{j}, \textit{RSSI(j,m)} is the RSSI
between node \textit{j} and the target node \textit{m}. Placing equation~\ref{eq3} into equation~\ref{AC} gives:

\begin{equation}
\label{ACOEx}
p_{ij}^k (t) = \left\{ \begin{array}{l}
\frac{  \tau_{ij}(t)^\alpha *  \{ 
{ max [RSSI(i, j ) + RSSI( j,m)] }  \}^\beta }
{\sum_{j}{[\tau_{ik}(t)]^\alpha * \{ 
{ max [RSSI(i, j ) + RSSI( j,m)] }  \}^\beta }  } 
 {\kern 11pt} , if j \in allowed_{k} \\
 0 {\kern 93pt}   , Otherwise \\
 \end{array} \right.\
\end{equation}

The pheromone trail updating mechanism helps to evaluate the quality of the solution and it consists generally of local and global updates. The local update is expressed in the following equation :

\begin{equation}
{\tau _{ij}(t+1)} = (1- \rho) \tau _{ij}(t) +  \Delta \tau _{ij} 
\end{equation}
where $\rho$ represents the evaporation of trail, \textit{t} is the iteration counter,  $\rho \in [0, 1]$ is the parameter that
regulates the reduction of $\tau _{ij}$ , and $\Delta \tau _{ij}$  is the value of pheromone in the current iteration which is deposited on the edges. $\Delta \tau _{ij}$  is expressed as follows:

\begin{equation}
{\Delta \tau_{ij}} = \sum_{k=1}^{M}\Delta\tau_{ij}^{k}
\end{equation}
where \textit{M} represents the total ants' number.

The pheromone amount deposited by ant \textit{k} traveling from the \textit{i}th node and coming at the \textit{j}th node is calculated as follows :
\begin{equation}
{\Delta \tau_{ij}^{k}} = 
\frac{Q}{L_{ij}}
\end{equation}
where \textit{Q} is a constant and \textit{Lij} is the distance traveled by the ant \textit{k} from the node \textit{i} to the node \textit{j}.  

In the traditional ACO and its subsequent versions concentrate on finding the shortest path while ignoring the effects of statistic metrics of energy parameter (e.g., average energy, minimum energy) which will lead to premature death of some sensor nodes and accordingly impact the lifetime of the overall network. In this paper, initial energy, average energy, minimum energy and path length are considered simultaneously. After the ants reach their destination, each ant represents a path. Here, we propose a fitness function conceived in such a way to select optimal route. Therefore, the fitness value of each path is computed as follows.

\begin{equation}
{fit_{m}^{k}} = 
\frac{E_{avg}*E_{min}}{e^{E_{init}}*L_{m}^{k}}
\end{equation}

where, $E_{avg}$, $E_{min}$, $E_{avg}$ and $E_{init}$ represent the average node energy, node minimal energy and node initial energy of ants passing on the path, and $L_{m}^{k}$ denotes the path length for \textit{mth} ant and \textit{kth} iteration.

It interesting to mention that, the optimality of the path is very related to the fitness value. That is, an optimal path is found when the fitness value is large. Then, the concentration of pheromone is updated on this path. Finally, the path with high fitness value is selected after several iterations and accordingly the balancing of energy consumption is improved. Generally, the balancing of energy impacts positively the entire network lifetime.

\begin{figure}[h]
\centering
\includegraphics[width=8.5cm]{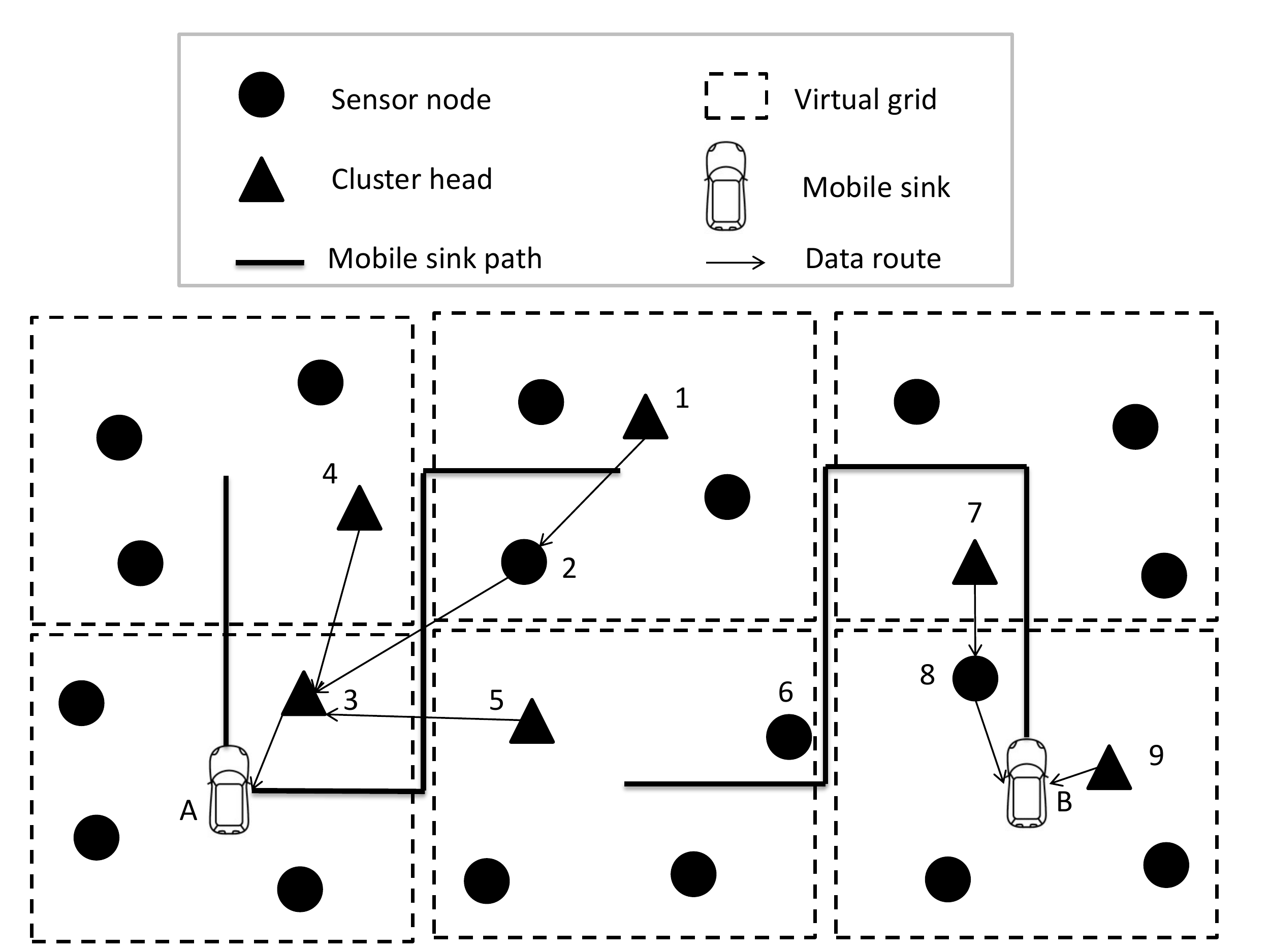}
\caption{Data routing with 2 mobile sinks.}
\label{fig33}
\end{figure}

For a better understanding of our multi-hop routing based on an improved ACO algorithm, we present the network model in figure~\ref{fig33} where we assume the mobile sink \textit{A} arrives in the virtual grid managed by CH 3. Hence, the CH 3 sends directly its data to the sink but the other CHs i.e. CH 1 and CH 4 needs to trigger the improved ACO algorithm to send their data. In such situation, the CH 1 sends its data to relay node 2 and this node sends data to CH 3 and then the data will arrive to sink. In the same way the CH 4 sends its data to the sink \textit{A} via the CH 3. Also the CH 5 sends its data  to sink \textit{A} through CH 3 and  CH 7 sends its data to sink \textit{B} through relay node 8 while CH 9 forwards directly its data to sink \textit{B}.

\begin{algorithm}
	\caption{Improved ACO algorithm for data transfer} 
 \hspace*{\algorithmicindent} \textbf{Input :}  Sensor nodes \\
 \hspace*{\algorithmicindent} \textbf{Output :}  The best route for data transfer

	\begin{algorithmic}[1]
		\For {$i=1$ to $lastant$}
			\State Calculate $RSSI(i, j )$ and  $RSSI( j,m)$
			\State Repeat until $ant(i)$ has completed a tour.
\State Select the node $j$ to be visited next.
		\EndFor
		\State Replacing higher-performance solutions.
\State Update the amount of pheromones.
\State The best solution is selected.
\State Transfer data packet from source to destination (sink).
	\end{algorithmic} 
\end{algorithm}

\section{Theoretical analysis}
\label{ther}
In this section, we analyse the delay and network lifetime of the proposed routing protocol.

\subsection{Delay analysis}
Many applications in event-driven WSNs are very critical to delay. Due to the dynamic environment, the long delay in the
delivery of events can cause serious and critical problems. To simplify, the delay can be defined by the hop counts 
from the node detecting the event to the sink. Delivering an information requires it to be traveled from the source node (S) to the sink node (SN) through a set of relay nodes (R). Thus, the \textit{k}th route of the packet \textit{i} is represented as follows.

\begin{equation}
R_{i,k} =  \left \{ (S_{i}),(R_{i}),(SN_{i})  \right \}
\end{equation}
where \textit{i} denotes the \textit{i}th packet.

To find the total route distance \textit{rD} traveled by the packet \textit{i}, we compute the following:

\begin{equation}
rD(R_{i,k}) = dis(S_{i}, R_{i,j})+
\sum_{j=1}^{n-1}dis(R_{i,j} , R_{i,j+1})
+ dis(R_{i,n} , SN_{i})
\end{equation}
where, dis(.,.) denotes the distance between the source and relay node or between relay nodes or between an \textit{n}th relay node and the sink.

The best solution is the one with the least route length $rD(R_{i,k})$. Therefore, to decide the best route, we have the following optimization problem.

\begin{equation}
min_{R_{i,k}}rD(R_{i,k})
\end{equation}

It is travelling salesman problem termed as TSP, which is NP-hard. To bypass this difficulty, we have proposed an improved ACO algorithm which tends to select the best route in terms of hop distance and accordingly the delay is enhanced at a high extent. In the state-of-the-art routing protocols based on mobile sinks, the data transfer is conditioned with the existence of the sink in a given cluster which  broadens the delay especially in large scale networks. Therefore, in order to reduce this delay, the improved ACO algorithm is applied to send immediately collected data and reach the sink where it is.

\subsection{Lifetime analysis}

The lifetime of the network is known as the duration
of the network until half of nodes run out of energy \cite{Ref383}. The half of nodes failure could lead to network partitioning
in some WSN systems and interruption of additional services. For this reason, this definition is very useful to evaluate the performance of any routing protocol. Overall, the
lifetime (\textit{L}) of any node \textit{i} with initial energy $\varepsilon_{ini}$ id defined as follows. 

\begin{equation}
Li=\frac{\varepsilon_{ini}}{e_{i}}, \forall i = 1,2,...,\eta 
\end{equation}
where $e_{i}$ is the average energy power per unit time.

We assumed that the energy is
consumed in the sending, receiving, encoding, idling, and listening modes. Consequently, the virtual network lifespan is then expressed as follows :
\begin{equation}
L^{Vir}=\frac{\eta \varepsilon_{ini}}{e{i}}
\end{equation} 

In this case, no energy is left in the network; however, in reality this is not the case since there exits some energy left at the end.  Consider a \textit{m} number of sensor nodes have remaining power $\varepsilon_{res,i}$, where $\eta$ is the total number of sensor nodes. The energy left in the network is seen as wasted energy. Thus, the average quantity of wasted energy
in WSN can be defined as follows :

\begin{equation}
\varepsilon_{w} =  \sum_{i=1}^{m}\varepsilon_{res,i}, t \geq lifetime
\end{equation}

Based on wasted energy $\varepsilon_{w}$, the network lifetime can be expressed as follows.

\begin{equation}
\label{eq24}
L=\frac{\eta  \varepsilon_{ini} - \varepsilon_{w}} {e_{i}}
\end{equation}

In equation~\ref{eq24}, $\eta$  and $\varepsilon_{ini}$ are constants; that is, the network lifetime is very related to $\varepsilon_{w}$. In \cite{Ref461}, it is stated that the network lifespan does not depend  only on the
average power consumed, but also on the leftover energy residual in the network.
Thus, the routing protocol should minimize the wasted energy
 $\varepsilon_{w}$ to a minimum
value min$\varepsilon_{w}$ in order to increase the network lifetime to a maximum value as max\textit{L} which is given by :

\begin{equation}
maxL=\frac{\eta  \varepsilon_{ini} - min\varepsilon_{w}} {e_{i}}
\end{equation}

Our proposed protocol not only selects higher energy nodes as CHs and performs re-clustering on the basis the CHs' energy but also it uses an improved ACO algorithm that favors higher energy paths to be selected for data delivery to the mobile sink. Therefore, this influences positively the balance of energy consumption among all of WSN nodes, which reduces the leftover wasted energy, and as a result the network lifetime is increased.
 
\section{Performance evaluation}
\label{perf}
In order to evaluate the performance of EHRP, extensive simulations are conducted on NS-2 simulator. The experiments are conducted to study the influence of EHRP on constrained network resources. The consequences are analyzed in terms of energy consumption, network lifetime, average delay, packet delivery ratio w.r.t. varying number of nodes, mobile sink speed, network dimension, and message size.  The
performance of EHRP is compared with those of two recent state-of-the-art routing
protocols namely IoGHR and ECMMS. Sensor nodes are
deployed randomly in the region. There
is a mobile sink node in the network and its movement path
is predefined. Table~\ref{val1} lists the parameter settings for the simulations.
To overcome the stochastic nature of WSN deployment,
each experiment is conducted 25 times and the results are reported as
their average.

For the ACO algorithm, intensive experiments were performed to define the appropriate values
for $\alpha$, $\beta$ and $\rho$ . As shown in table~\ref{val}, when $\alpha$, $\beta$ and $\rho$ have values of 5, 10 and
0.6, the path length is the smallest and the number of iterations is also small.

\begin{table}
\caption{Simulations parameters. }
\centering
\begin{tabular}{p{4cm} p{3cm}  }
\hline\noalign{\smallskip}

Parameter & value \\
\noalign{\smallskip}\hline\noalign{\smallskip}
Control packet size & 200 bits \\
Medium access control (MAC) protocol & Tunable MAC \\
Initial energy & 0.5 J \\
$E_{fs}$ & 10 pJ/bit/$m^{2}$ \\
$E_{mp}$ & 0.0013 pJ/bit/$m^{4}$ \\
$E_{elec}$ & 50 nJ/bit \\
$E_{DA}$ & 5 nJ/bit \\
$d_{0}$ & 87.705 m \\
 \noalign{\smallskip}\hline
\end{tabular}
\label{val1}
\end{table}

\begin{table}
\caption{Length of path by using different values for $\alpha$, $\beta$ and $\rho$. }
\centering
\begin{tabular}{p{0.7cm} p{0.7cm} p{0.7cm} p{2cm} p{2cm} }
\hline\noalign{\smallskip}

$\alpha$ & $\beta$ & $\rho$ & Path length & Iteration \\
\noalign{\smallskip}\hline\noalign{\smallskip}
0.5                   & 1                    & 0.3                 & 638.02      & 78        \\ 
0.5                   & 1                    & 0.4                 & 639.33      & 87        \\ 
0.5                   & 1                    & 0.5                 & 639.33      & 95        \\ 
0.5                   & 1                    & 0.6                 & 639.59      & 92        \\ 
1                     & 5                    & 0.3                 & 636.95      & 67        \\ 
2                     & 6                    & 0.4                 & 637.29      & 51        \\ 
4                     & 8                    & 0.5                 & 637.13      & 58        \\ 
5                     & 10                   & 0.6                 & 636.74      & 59        \\ \noalign{\smallskip}\hline
\end{tabular}
\label{val}
\end{table}

In the following, we will present the performance metrics as well as the simulation results and discussions.

\subsection{Performance metrics}
For evaluating the efficacy of EHRP, the following performance metrics are considered:
\begin{itemize}
\renewcommand\labelitemi{--}
\item Network lifetime: Defined as the time of
half of the sensor nodes failure.
While in some literature the time before the first sensor node dies is known as the
network's lifetime, the period when the first node dies can not mean the network's lifetime is over.  If the node at the edge of the network fails, the remaining nodes will still operate well
and, from the point of view of its functionality, the network performance can not be affected roughly. At this point, it is reasonably believed
that the lifetime of the network will continue \cite{Ref383}.
\item Energy consumption of the network : It is the overall energy consumed by sensor nodes.
\item Average delay : It is the average
delay of all received data packets from destination. The smaller the value is,
the better the performance of the routing protocol. 
\item Packet delivery ratio : Is the ratio of data packets successfully collected by the sink to the total number of transmitted data packets.
\end{itemize}

\subsection{Simulation results and discussions}
In the following, we will see the influence of the number of sensor nodes, the sink speed, the network dimension and the message size on the performance of the EHRP, IoGHR and ECMMS.

\subsubsection{Impact of number of sensor nodes}

In this section, EHRP is evaluated according to the aforementioned performance metrics and compared with ECMMS and IoGHR to observe the effect of number of nodes. However, the message size is set to 4000 bits, the network dimension is set to 1000*1000 $m^{2}$, and the speed of the sink is kept fixed at 9 km/h.

\begin{figure*}[!h]
\begin{subfigure}{0.48\linewidth}
    \includegraphics[width=\linewidth, height=0.7\linewidth]{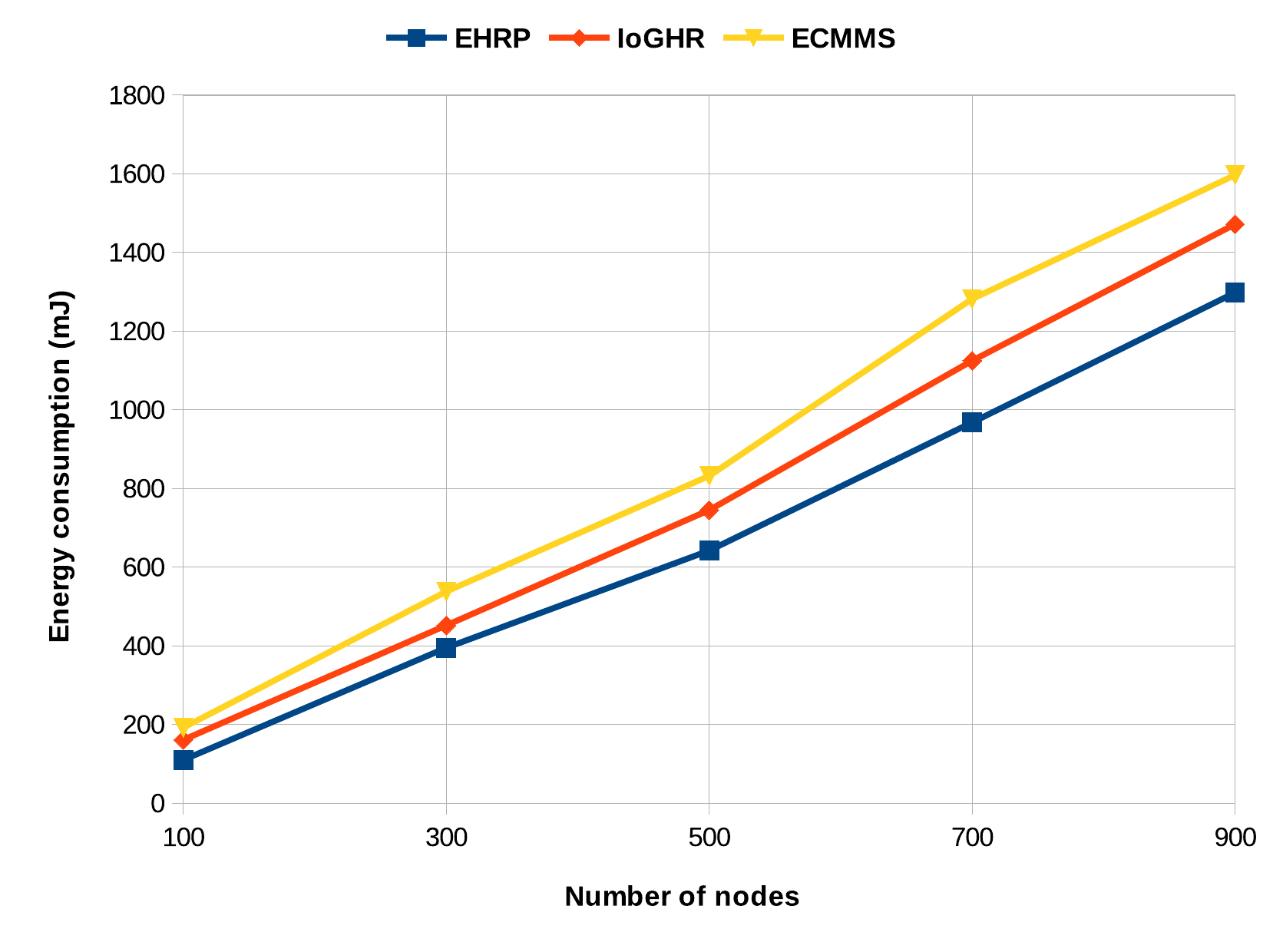}
\caption{Total energy consumption.}
    \label{ener}
\end{subfigure}
    \hfill
\begin{subfigure}{0.48\linewidth}
    \includegraphics[width=\linewidth, height=0.7\linewidth]{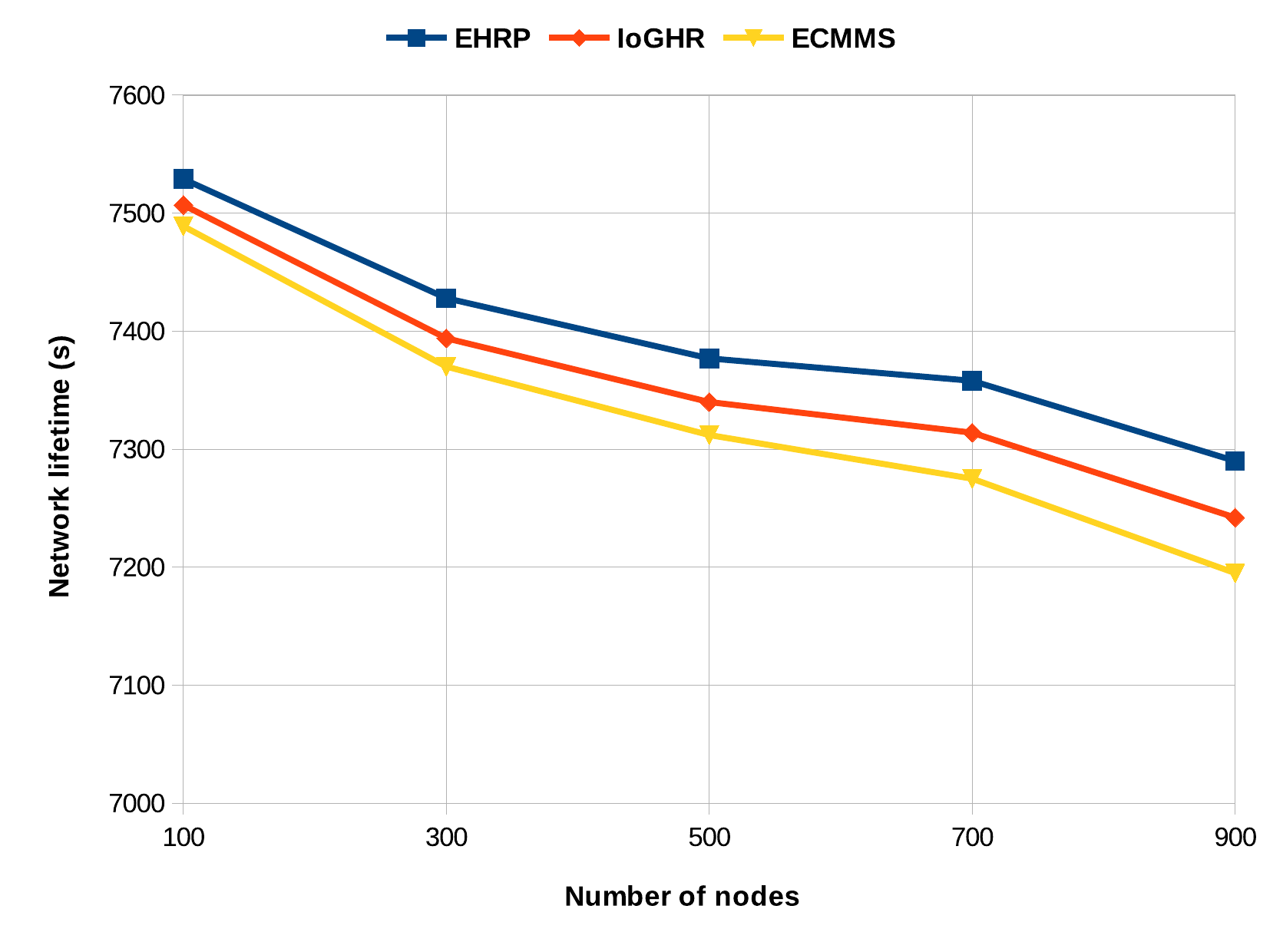}
\caption{Network lifetime.}
    \label{life}
\end{subfigure}

\begin{subfigure}{0.48\linewidth}
    \includegraphics[width=\linewidth, height=0.7\linewidth]{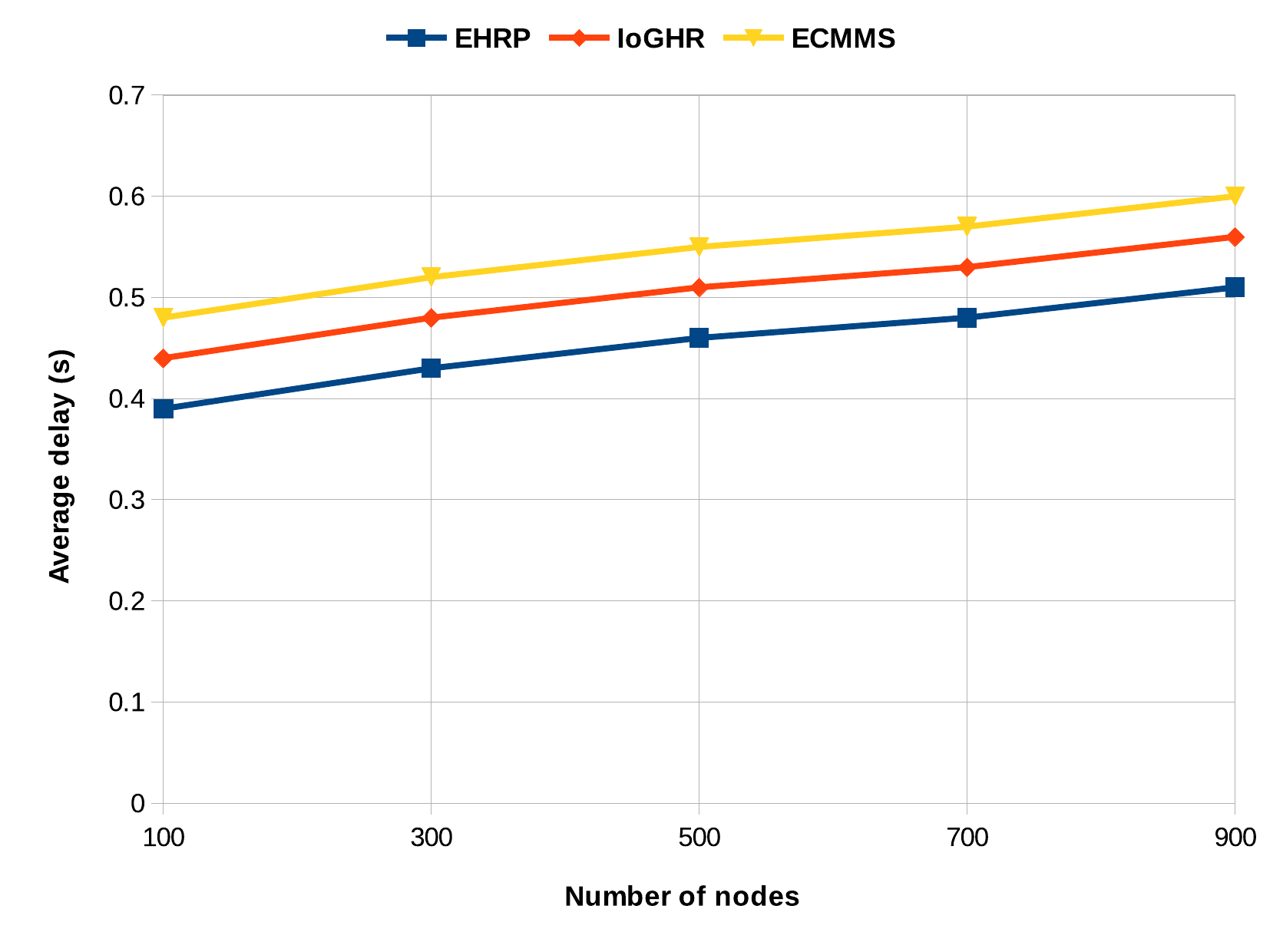}
\caption{Average delay.}
    \label{del}
\end{subfigure}
    \hfill
\begin{subfigure}{0.48\linewidth}
    \includegraphics[width=\linewidth, height=0.7\linewidth]{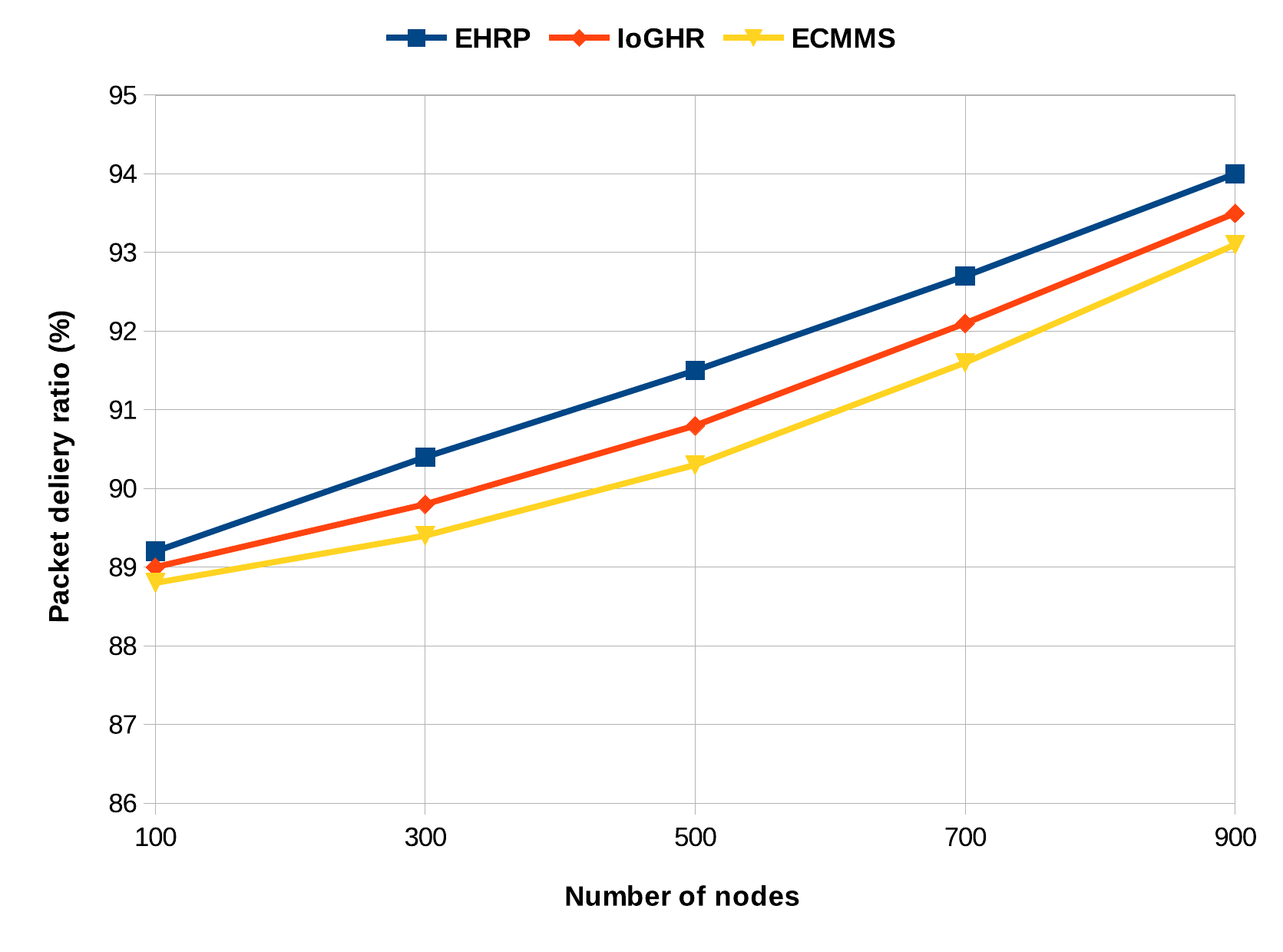}
\caption{Packet delivery ratio.}
    \label{pdr}
\end{subfigure}
\caption{Impact of number of sensor nodes on network performance.}
\end{figure*}

\begin{alphalist}
    \item Energy consumption and network lifetime : The energy consumption and the network lifetime of EHRP,
ECMMS and IoGHR for varying the number of nodes are observed in figure~\ref{ener} and figure~\ref{life} respectively. The energy consumption shows growing pattern  for all the
three routing protocols as the number of sensor nodes in the network
increases. This is due to the fact that with the increase in number of nodes,
the number of nodes encountered when transmitting the data to sink increases. However, EHRP results in low  energy consumption
as compared to ECMMS and IoGHR. This is mainly due to the rotation of CH role in the round-robin method based on energy which is ensured in EHRP in contrast to the other protocols that do not use it. This method allows to save more energy because the network setup is done only once until the end of network lifetime which avoids repetitive control overhead required to cluster the network. Also, the fitness function in the proposed ACO algorithm considers average energy, minimal energy and route length. Thus, if the fitness value is low, the updated pheromones' quantity will be low. Accordingly, after several iterations, the path will be forsaken by ants and this enhances the energy consumption balancing which protracts the network lifetime. Additionally, the heuristic function is conceived to save energy by choosing higher energy nodes and avoiding retransmissions.  IoGHR performs better than ECMMS in energy consumption metric as noticed in the figure and this is due to the fact that ECMMS requires to exchange more overhead to create partitions unlike IoGHR that does not require this mechanism and thus less overhead is used. The energy consumption has a direct impact on the network lifetime, and this justifies the increased network lifetime of our proposed protocol in comparison with IoGHR and ECMMS respectively.

    \item Average delay : Figure~\ref{del} depicts the average delay of
EHRP, ECMMS, and IoGHR for the varying the number of nodes. This metric increases with respect to the number of deployed nodes. As the
the number of nodes in the network evolves, the number of member nodes per cluster increases, which further causes more delay to gather all the data before combining them into a single message and forwarding them consequently. However, compared
to ECMMS and IoGHR, the average delay of EHRP is much
lower due to shortest data delivery routes adopted using together our improved ACO based routing algorithm and mobile sink technology. This allows CHs to avoid waiting for mobile sink arrival to send data (the case of ECMMS and IoGHR) but rather the data are sent via the most shortest route. Also, we can notice that IoGHR presents an advantage in comparison to ECMMS according to average delay and this is due to the regularity of distance between grids in IoGHR contrariwise to ECMMS which presents some irregularities in distances between clusters.

    \item Packet delivery ratio : The performance of EHRP concerning
data delivery ratio while increasing the number of nodes is shown and compared with that of ECMMS and IoGHR in figure~\ref{pdr}. The data delivery ratio
increases with respect to increase in the number of deployed sensor
nodes in all the three routing protocols. This is due to the fact that the number
of source nodes evolves with the increase of sensor
nodes deployed in the same area. It therefore leads to increase the number of data packets received at the sink which resulting in a higher data transmission ratio. Unlike ECMMS and IoGHR, EHRP performs well with respect to the number of sensor nodes. This expected behavior results from the fact that EHRP is more energy efficient and reliable in comparison with other protocols. Also, it can be noticed that IoGHR performs better in comparison with ECMMS due to the short distances between virtual grid which means shortest path for mobile sink which increases the frequency of data gathering in comparison with that of ECMMS.

\end{alphalist}

\subsubsection{Impact of sink speed}
In this section, EHRP is evaluated concerning the aforementioned
performance metrics and compared with ECMMS and IoGHR to
observe the effect of varying sink speed. However, the number of sensor nodes in the network is kept fixed at 900, message size is set to 4000 bits, and the network dimension is set to 1000*1000 $m^{2}$.

\begin{alphalist}
    \item Energy consumption and network lifetime : The energy consumption and the network lifetime of EHRP,
ECMMS, and IoGHR at various sink speeds are illustrated in figure~\ref{ener1} and figure~\ref{life1} respectively. The energy consumption plot
 shows rising
pattern in function of the increment in the
sink speed. This is due to the fact that the more the speed of the mobile sink increases the more the CHs are solicited and hence more communications are done between nodes. From the results, it is clear that the sink speed affects negatively the energy consumption and network lifetime. The EHRP and IoGHR show better performance in terms of energy consumption and network lifetime against ECMMS owing to the fact that these routing protocols do not require any sink presence announcement when it comes within a cluster unlike ECMMS where the sink has to declare its presence every time. The better performance of EHRP over IoGHR is due mainly to the lower resulting overhead for network clustering.

  \item Average delay : Figure~\ref{del1} shows the average 
delay of EHRP when compared to ECMMS and IoGHR for varying
sink speeds. In the case of high sink speed, the sink moves in a
short span of time over a fairly large portion of the network. consequently, gathering the
data will take less time and the data packets could be
sent to the sink with less delay.  It is also noticed that EHRP presents better performance in comparison with other protocols and this is due to the ACO-based multi-hop routing method selecting short routes which supports the advantage of high speed of mobile sink. As noticed in the figure, IoGHR has better performance with respect to ECMMS because of the sink solicitation of the CH to inform it about its presence. With the increase in sink speed more solicitation happened and then the data delivery is delayed.

  \item Packet delivery ratio : Figure~\ref{pdr1} shows the data delivery ratio of
EHRP, ECMMS and IoGHR at different sink's speeds.
In the case of high sink speed , in a short time the sink
moves in a large portion of the network and it would
collect a high number of data packets. This justifies the increase in packet delivery ratio. EHRP outperforms its peers due to the data delivery policy in which CHs do not wait for mobile sink coming but they continue to transfer data to sink. Also, IoGHR presents better data delivery ratio in comparison with ECMMS because of the time schedule of the mobile sink to CH nodes which avoids extra wait-time as in ECMMS and also the behavior is justified by the better performance regarding the network lifetime. This means more nodes are alive and continue to send data to its destination.

\begin{figure*}[!h]
\begin{subfigure}{0.48\linewidth}
    \includegraphics[width=\linewidth, height=0.7\linewidth]{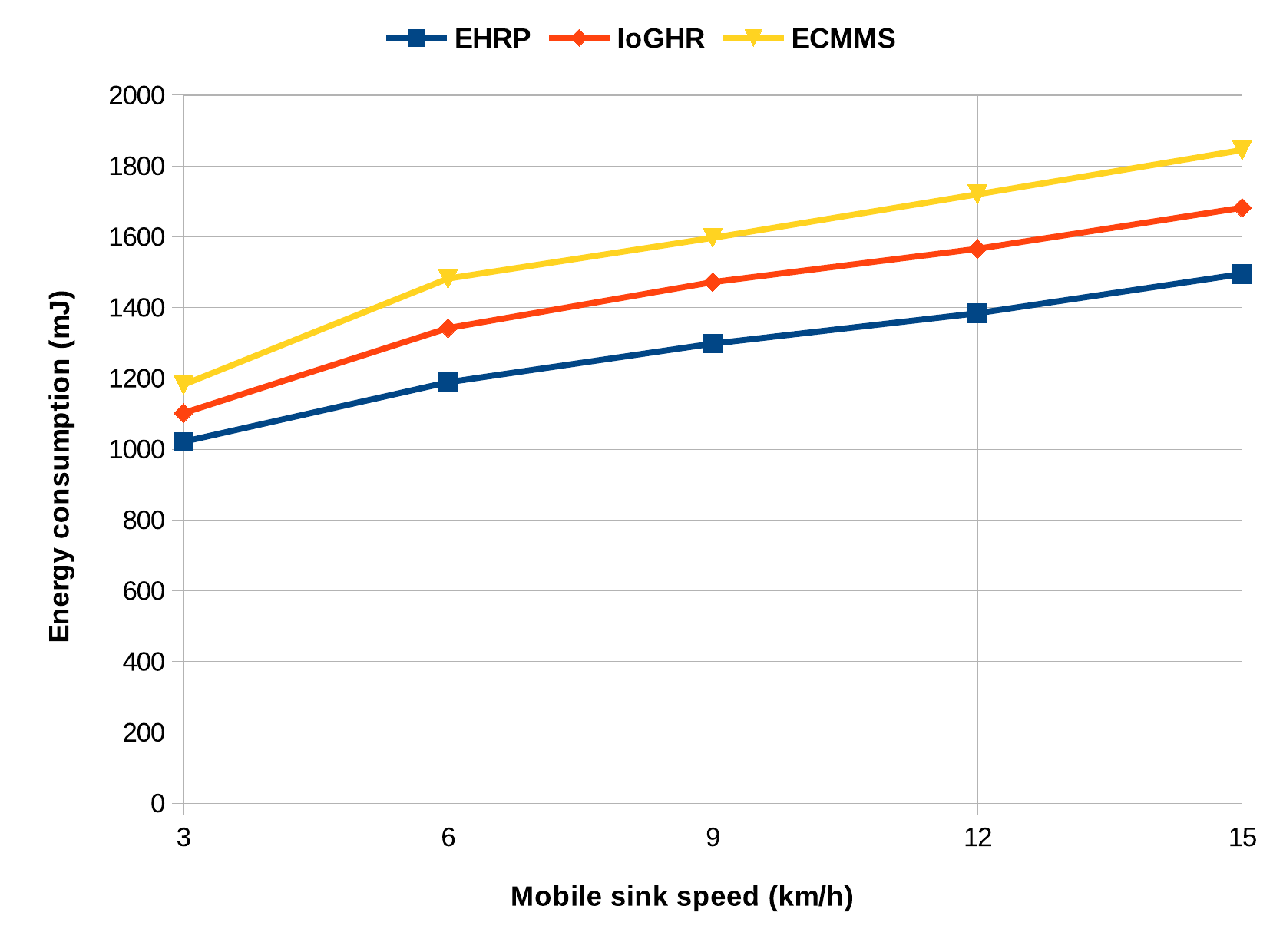}
\caption{Total energy consumption.}
    \label{ener1}
\end{subfigure}
    \hfill
\begin{subfigure}{0.48\linewidth}
    \includegraphics[width=\linewidth, height=0.7\linewidth]{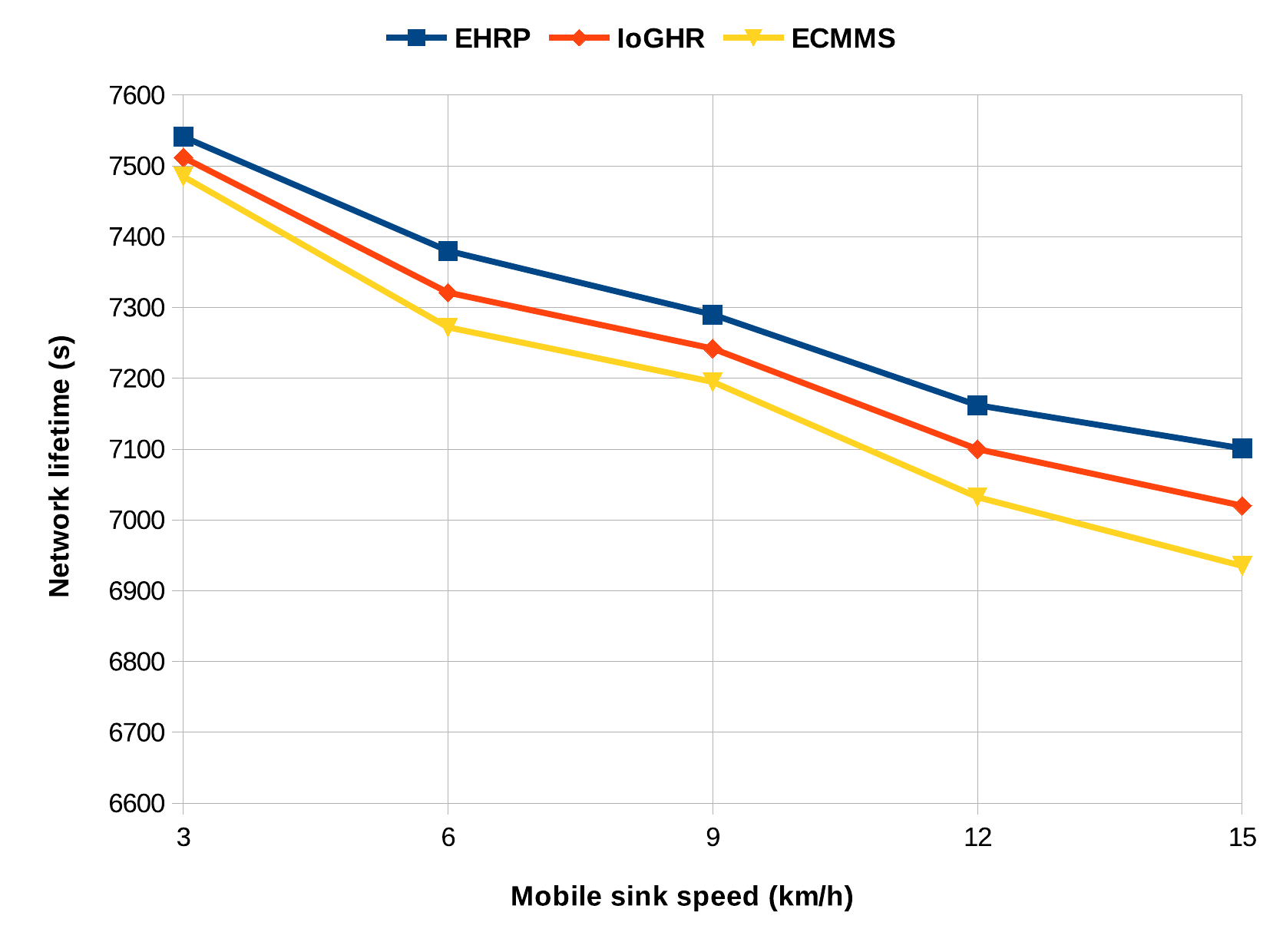}
\caption{Network lifetime.}
    \label{life1}
\end{subfigure}

\begin{subfigure}{0.48\linewidth}
    \includegraphics[width=\linewidth, height=0.7\linewidth]{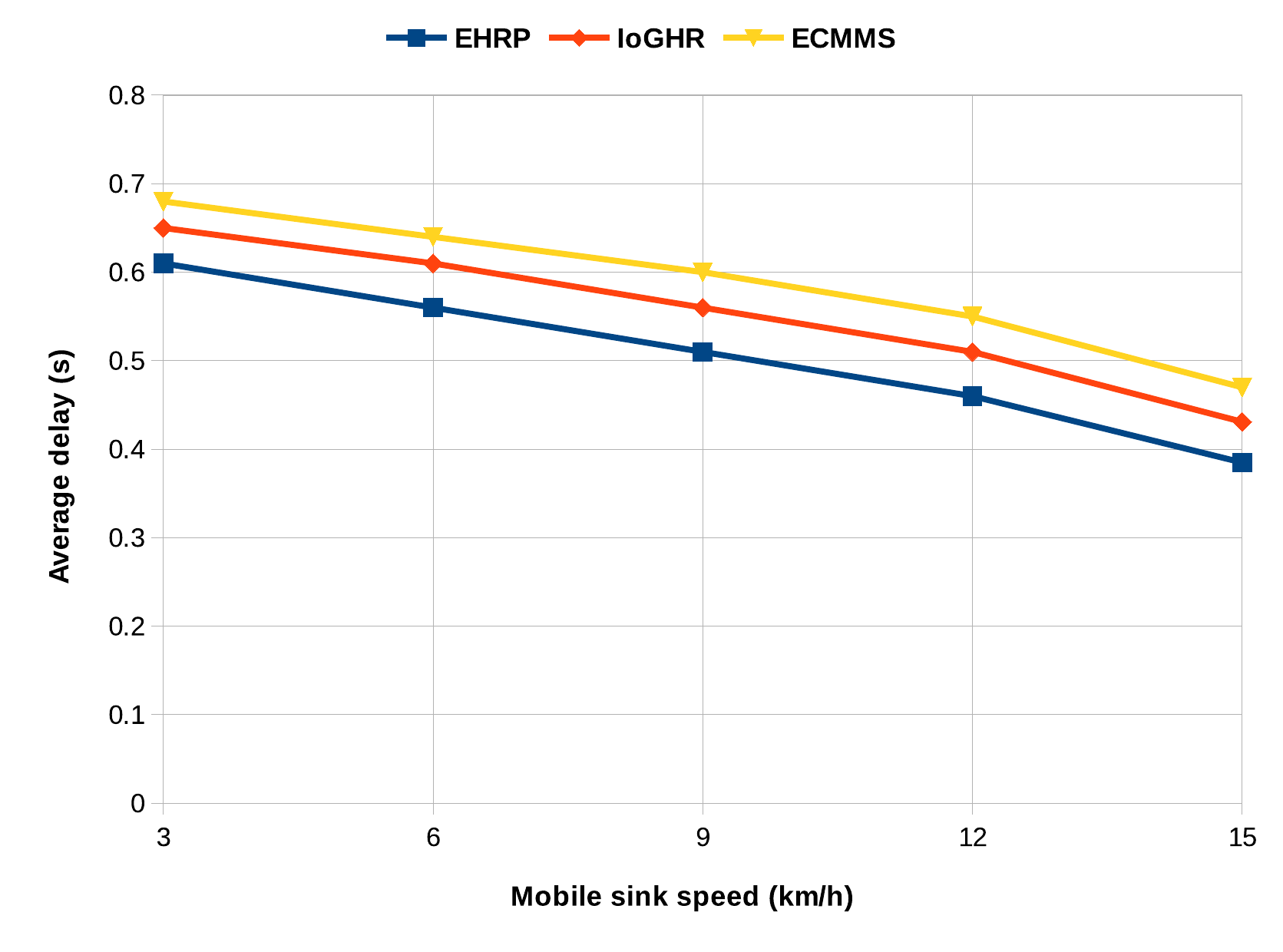}
\caption{Average delay.}
    \label{del1}
\end{subfigure}
    \hfill
\begin{subfigure}{0.48\linewidth}
    \includegraphics[width=\linewidth, height=0.7\linewidth]{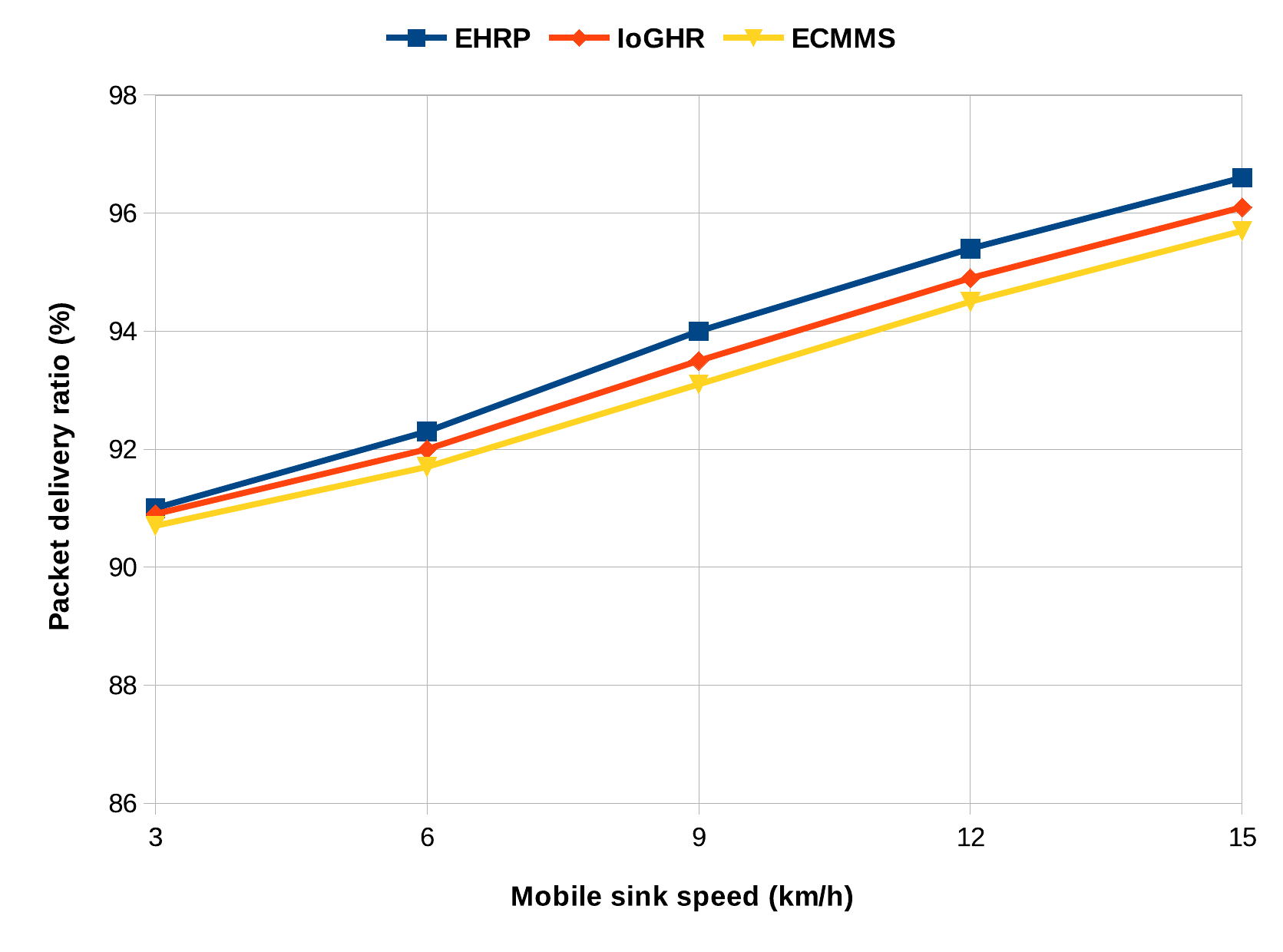}
\caption{Packet delivery ratio.}
    \label{pdr1}
\end{subfigure}
\caption{Impact of sink speed on network performance.}
\end{figure*}

\end{alphalist}

\subsubsection{Impact of network dimension}
In this section, EHRP is evaluated concerning the aforementioned
performance metrics by varying the network dimension and compared
with ECMMS and IoGHR. However, the number of sensor nodes, message size (in bits), and sink
speed (in km/h) are kept fixed at 900, 4000 and 9 respectively for the experiments.

\begin{figure*}[!h]
\begin{subfigure}{0.48\linewidth}
    \includegraphics[width=\linewidth, height=0.7\linewidth]{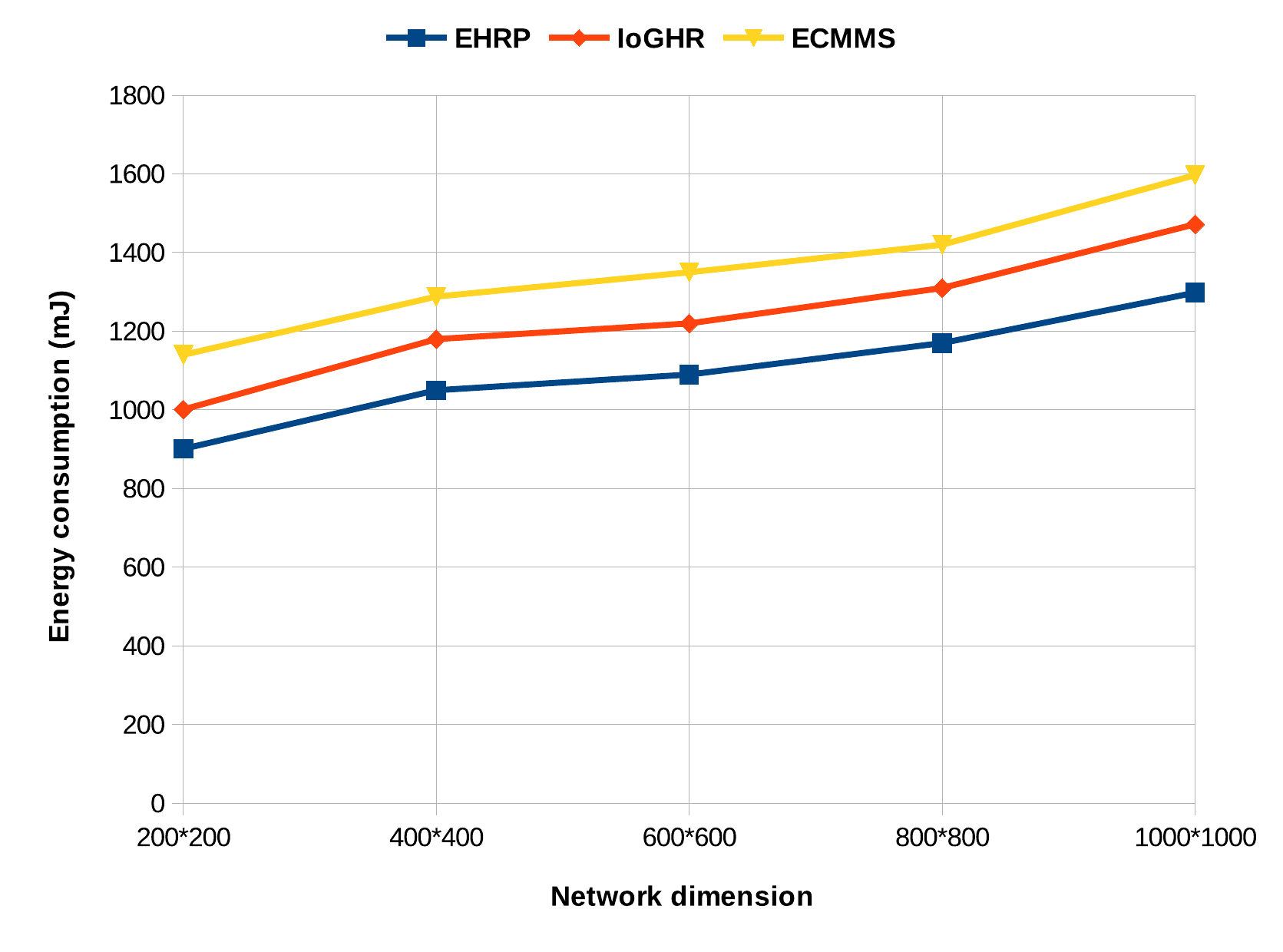}
\caption{Total energy consumption.}
    \label{ener2}
\end{subfigure}
    \hfill
\begin{subfigure}{0.48\linewidth}
    \includegraphics[width=\linewidth, height=0.7\linewidth]{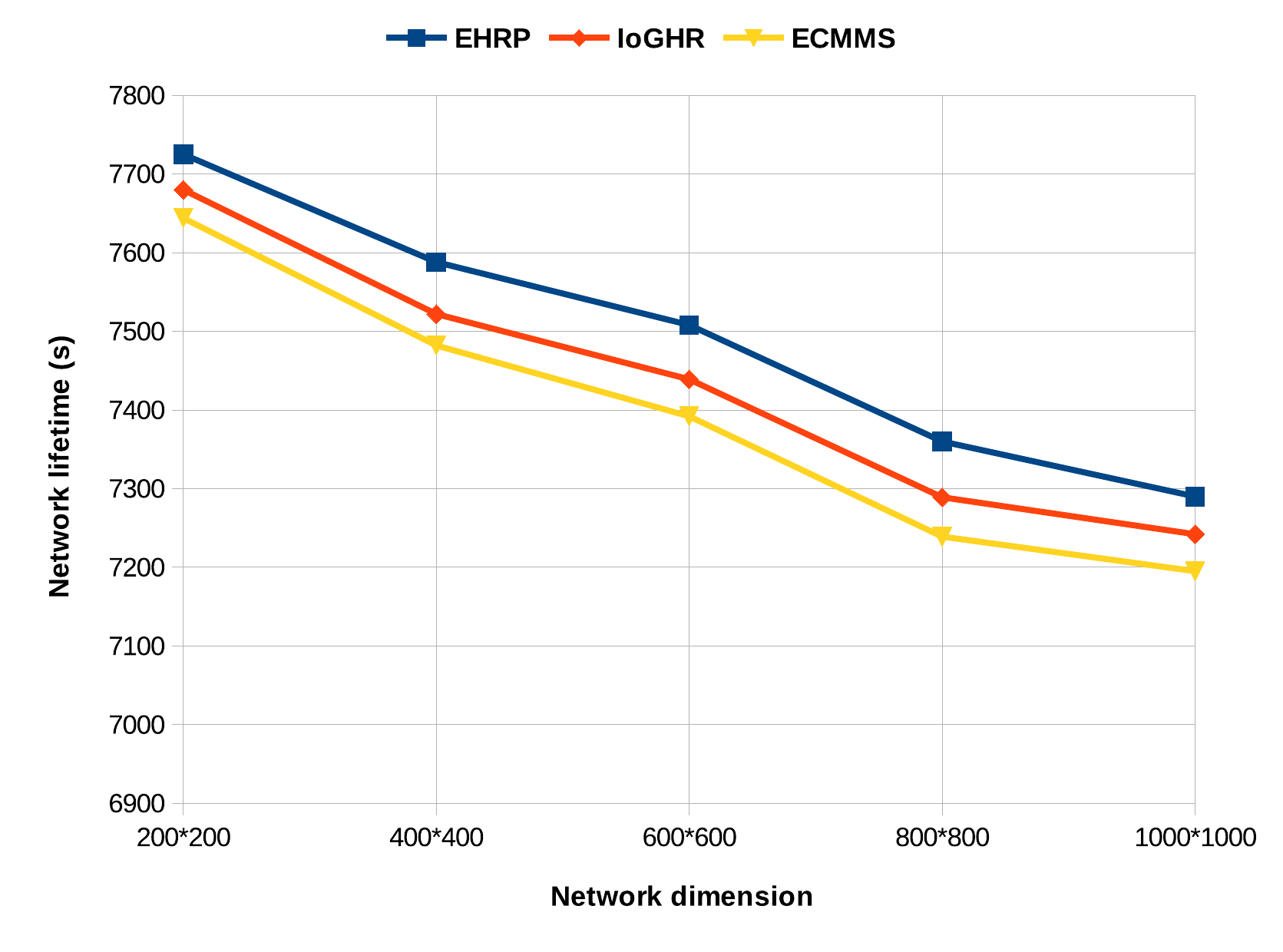}
\caption{Network lifetime.}
    \label{life2}
\end{subfigure}
\begin{subfigure}{0.48\linewidth}
    \includegraphics[width=\linewidth, height=0.7\linewidth]{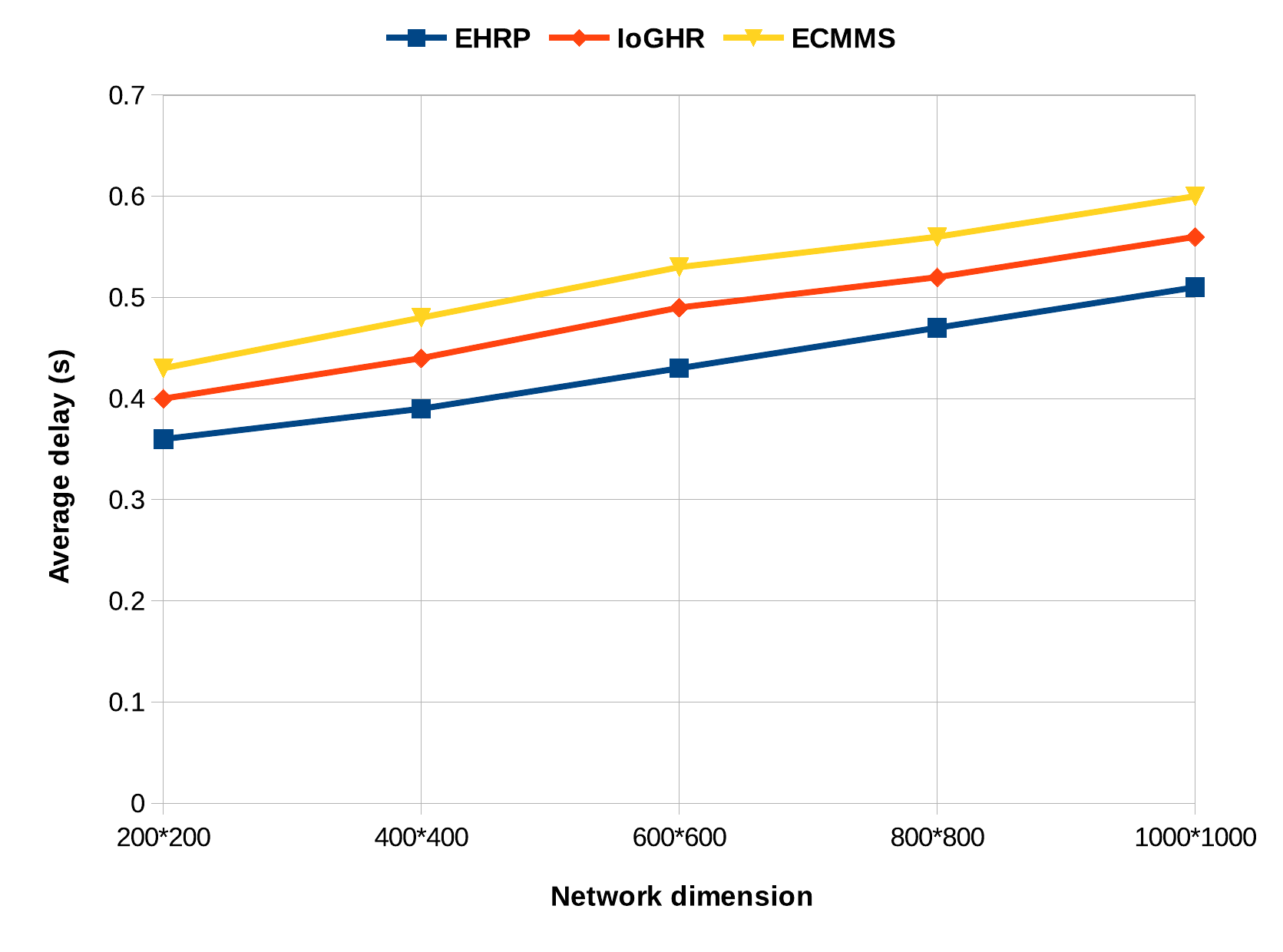}
\caption{Average delay.}
    \label{del2}
\end{subfigure}
    \hfill
\begin{subfigure}{0.48\linewidth}
    \includegraphics[width=\linewidth, height=0.7\linewidth]{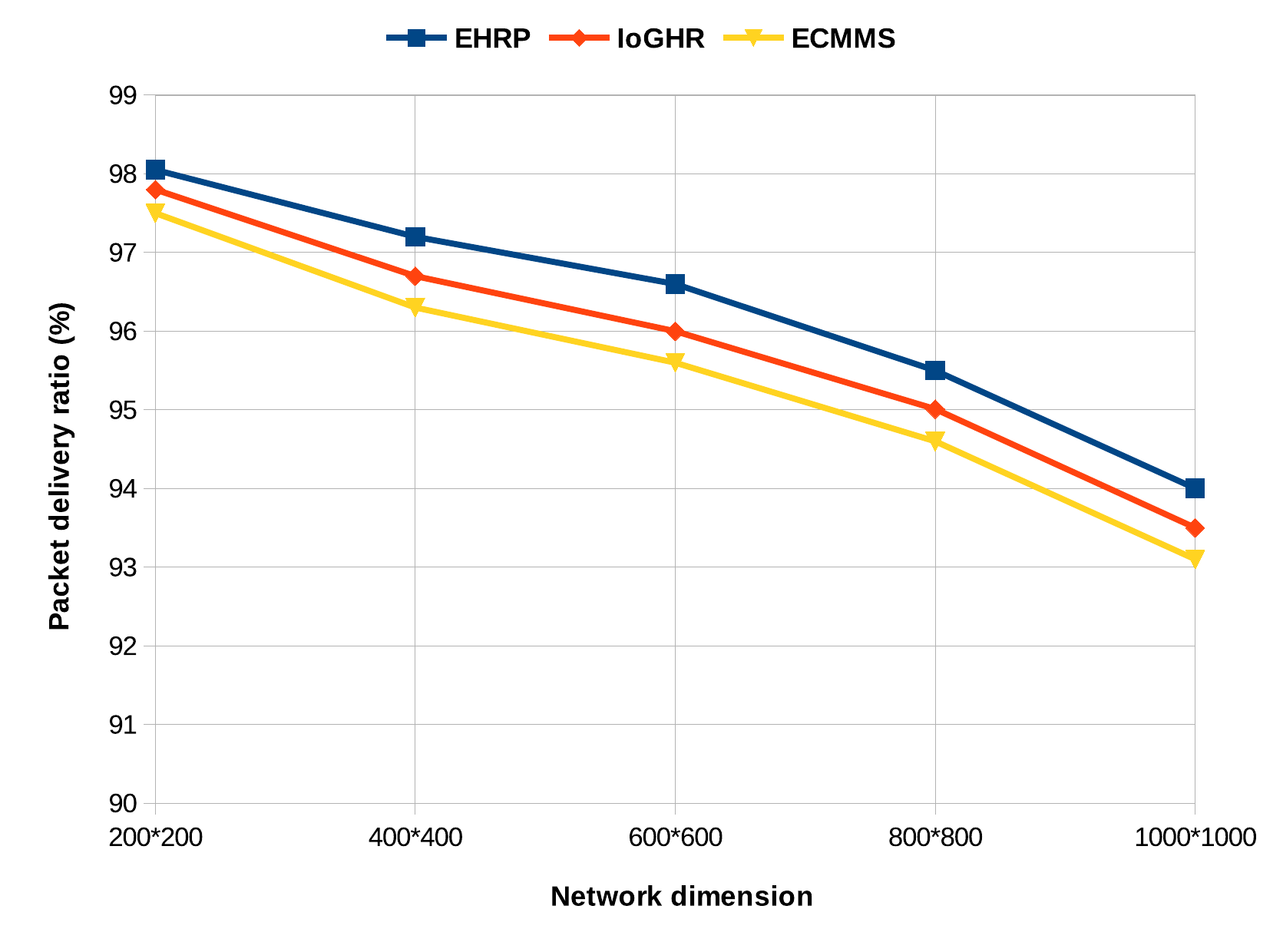}
\caption{Packet delivery ratio.}
    \label{pdr2}
\end{subfigure}
\caption{Impact of network dimension on network performance.}
\end{figure*}

\begin{alphalist}
    \item Energy consumption and network lifetime : Figure~\ref{ener2} and figure~\ref{life2} show the energy consumption and the network lifetime
of EHRP, ECMMS, and IoGHR for different network dimensions. The increment in network dimension
 increases also the distance required to transmit data among sensor nodes to the sink, which results in increased energy consumption and decreases consequently the network lifetime. The better performance of the proposed EHRP against ECMMS and IoGHR is due to the justification mentioned above and additionally the ECMMS routing protocol does not consider the connectivity which is a vital issue in large scale WSNs as some sensor nodes may not be attached to a CH which may result in excessive energy consumption because of idle listening of the rest of these unconnected nodes. IoGHR fixes the position of sensor nodes within a cluster and employs two-hop communication mode between the members of a cluster and the sink. As the distance affects the energy consumption, this says that the position of sensor nodes is predetermined, therefore they may be distant for the CH contrariwise to our mechanism of cluster formation that attaches the sensor nodes based on received signal intensity to the CHs and then the distance between these nodes is optimized. 
 
    \item Average delay : Figure~\ref{del2} depicts the average delay
of EHRP, ECMMS, and IoGHR for different network dimensions. With the increase in
network dimension, the distance for data transmission increases which hence increases delay. The expected better performance of EHRP against its peers comes mainly from the point of avoiding waiting-time of CHs to transfer data. Additionally, the better performance of EHRP and IoGHR over ECMMS is justified to the regularity of the deployment based principally on virtual grids which seems to have a marginal but existing effect on the network
performance i.e. delay, as the topological irregularities increase mobile sinks routes lengths due to irregularities in distances. 

    \item Packet delivery ratio : Figure~\ref{pdr2} depicts the packet delivery ratio of
EHRP when compared to ECMMS and IoGHR for different network dimensions. The largest sensor field size causes a degradation in data delivery ratio due to increased distance; however, the proposed routing protocol can satisfy the required packet
delivery ratio due to the single and multi-hop communication scheme implemented in our proposal. It is reliable as it is based on RSSI which tends to avoid the probability of sending failed data packets. This allows to send more packets and avoids the waiting-time presented in ECMMS and IoGHR and hence it increases eventually the packet delivery ratio. Although the communication in ECMMS and IoGHR is of two hops from the sensor node to the sink, still the topological irregularity in ECMMS leads to an increase in energy consumption due to irregularities in distances. Consequently, as the data transmission is related to the lifetime of sensor nodes, the more alive nodes exist, the more the transmission persists and accordingly the the data packet still delivered to the sink.
\end{alphalist}

\subsubsection{Impact of message size}
In this section, EHRP is evaluated concerning the energy consumption and network lifetime
and compared with the ECMMS and IoGHR to
observe the effect of varying message size. However, the number of sensor nodes in the network is kept fixed at 900, and mobile sink speed is fixed to 9 km/h, and network dimension is set to 1000*1000 $m^{2}$.

The size of the messages sent from the nodes to the CH and from the CH to the sink is constant and equal to 4000 bits in all experiments which have been done so far.  In the energy consumption model, the size of the messages is efficient for sending and receiving messages; thus, the greater the size
of the messages sent and received in the sensor nodes, the more power is consumed and the lifetime of the network is reduced. Through this experiment, we would like to show the influence of the size of
messages received and sent from sensor nodes on the performance of the comparative routing protocols. 

\begin{figure*}[ht]
\begin{subfigure}{0.48\linewidth}
    \includegraphics[width=\linewidth, height=0.7\linewidth]{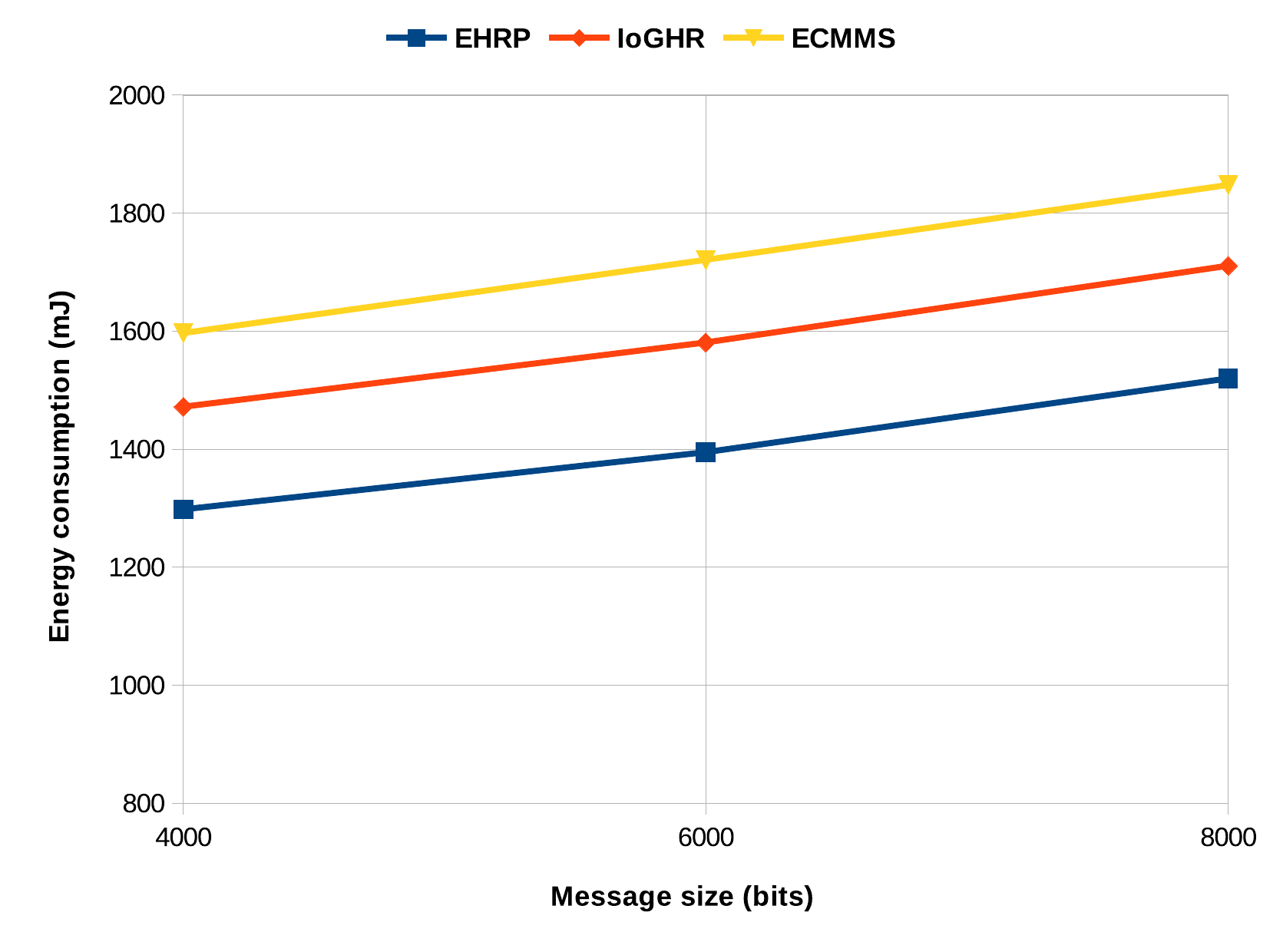}
\caption{Total energy consumption.}
    \label{ener3}
\end{subfigure}
    \hfill
\begin{subfigure}{0.48\linewidth}
    \includegraphics[width=\linewidth, height=0.7\linewidth]{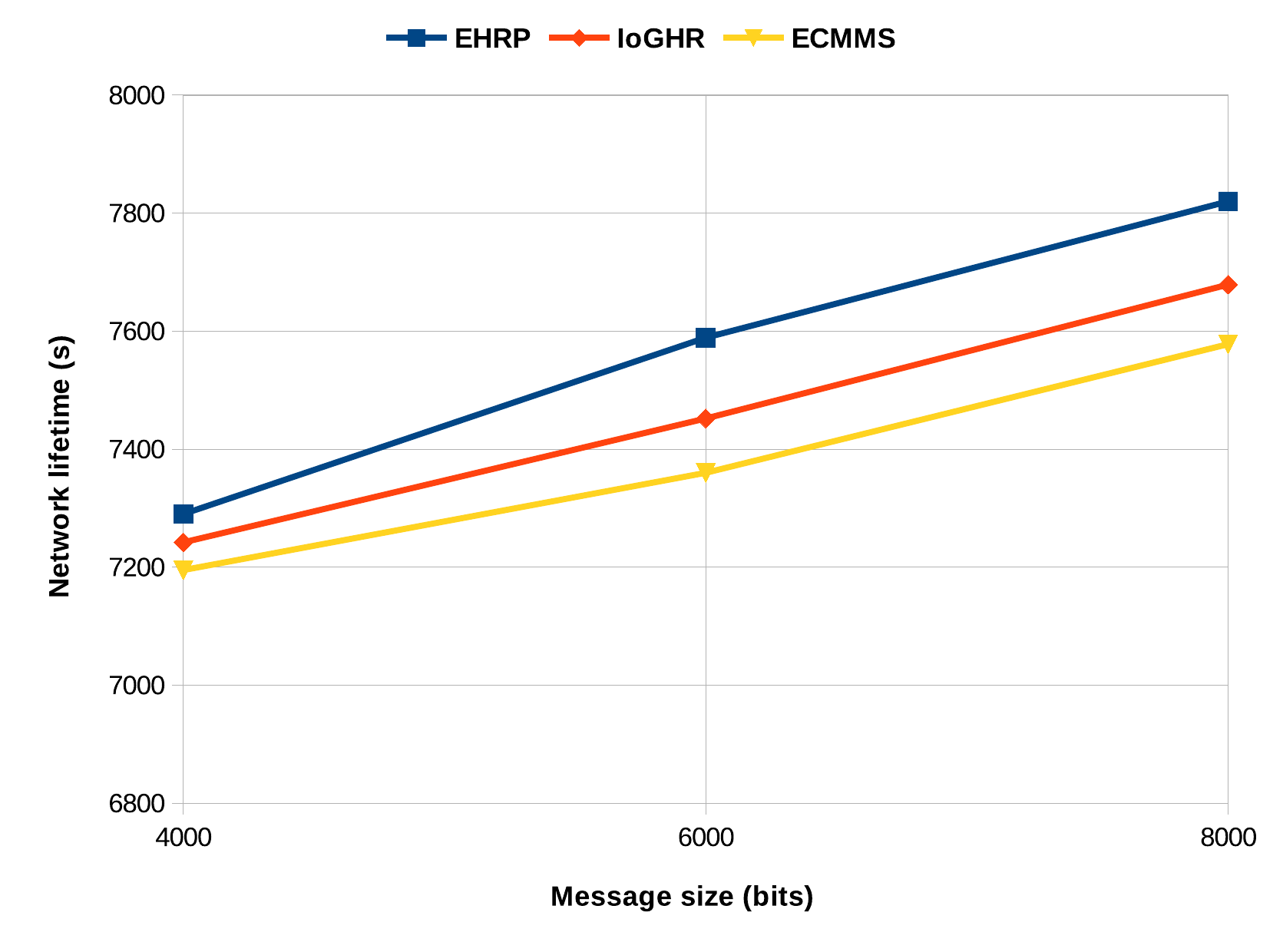}
\caption{Network lifetime.}
    \label{life3}
\end{subfigure}

\caption{Impact of message size on network performance.}
\end{figure*}

Figure~\ref{ener3} and figure~\ref{life3} depict the energy consumption and network lifetime of EHRP, ECMMS and IoGHR routing protocols. As previously
mentioned, cluster update in EHRP is only performed only once until the network service is ended, but in the
other routing protocols, clustering is performed in each round. The messages exchanged in each round to elect CHs and create clusters are greater; therefore, according to equation~\ref{eqq}, by increasing
the size and number of messages sent and received, more energy is consumed, and the lifetime of the network decreases. Additionally, less required overhead in IoGHR as compared to ECMMS leads to the better energy efficiency. 

\section{Conclusion and future work}
\label{conc}
Owing to the fact that existing state-of-the-art routing protocols fail to consider the compromise between energy efficiency and delay, this paper presents a routing protocol named EHRP to fix this issue. In this paper, we introduce an energy-efficient clustering method which results in having multiple rounds without any need for control overhead. Also, we proposed also an RSSI-based heuristic function for reliable data transmission and energy efficiency. Also, a fitness function was proposed to select optimal path while ensuring energy balancing. The major finding of our protocol is its exceptional performance over three of the most relevant works, namely, ECMMS and IoGHR in terms of energy consumption, network lifetime, average delay, and packet delivery ratio. This is justified by means of simulations of our proposal along with ECMMS and IoGHR. 

As a future work, we plan to consider the fault tolerant aspect of the mobile sink as it is not studied in literature. Therefore, the reliability of the proposed routing protocol will be enhanced. The future work should ensure the trade-off between reliability, energy consumption and delay for a better network performance.

\textbf{Funding} The authors received no funding for this research work.

\section{Ethics declarations}
\textbf{Conflict of interest} The authors declare that they have no conflict of
interest.


%
%

\bibliographystyle{spbasic}  
\bibliography{EHRP}
%
%

\end{document}